\documentclass[epjST]{svjour}
\usepackage{graphics}
\usepackage{graphicx}
\usepackage{amsfonts}
\usepackage{color}

\newcommand{\ssmall}[1]{\mbox{\footnotesize{#1}}}

\begin{document}
\title{Partial Transfer Entropy on Rank Vectors}
\author{Dimitris Kugiumtzis\thanks{\email{dkugiu@gen.auth.gr}}}
\institute{Department of Mathematical, Physical and Computational Sciences, Faculty of Engineering, Aristotle University of Thessaloniki, Thessaloniki 54124, Greece}
\abstract{For the evaluation of information flow in bivariate time series, information measures have been employed, such as the transfer entropy (TE), the symbolic transfer entropy (STE), defined similarly to TE but on the ranks of the components of the reconstructed vectors, and the transfer entropy on rank vectors (TERV), similar to STE but forming the ranks for the future samples of the response system with regard to the current reconstructed vector. Here we extend TERV for multivariate time series, and account for the presence of confounding variables, called partial transfer entropy on ranks (PTERV). We investigate the asymptotic properties of PTERV, and also partial STE (PSTE), construct parametric significance tests under approximations with Gaussian and gamma null distributions, and show that the parametric tests cannot achieve the power of the randomization test using time-shifted surrogates. Using simulations on known coupled dynamical systems and applying parametric and randomization significance tests, we show that PTERV performs better than PSTE but worse than the partial transfer entropy (PTE). However, PTERV, unlike PTE, is robust to the presence of drifts in the time series and it is also not affected by the level of detrending. 
} 
\maketitle
%
\section{Introduction}
\label{sec:intro}

A fundamental concept when studying coupled systems, and dynamical systems in general, is the dependence of one variable $X$ measured over time on another variable $Y$ measured synchronously, where both $X$ and $Y$ may be considered as variables of one single system or representative variables of coupled subsystems. The best known concept of interdependence in time series is Granger causality, defined in terms of predictability of linear stochastic systems \cite{Granger69}, i.e. $X$ Granger causes $Y$ if it improves the prediction of $Y$ when included in the autoregressive model. Granger causality has been quantified with measures based on linear models in the time and frequency domain \cite{Geweke82,Baccala01,Faes12b}. Besides the direct extension of Granger causality to nonlinear prediction models \cite{Schiff96}, the idea of Granger causality has been formulated focusing on specific system dynamics properties, ranging from point distance interdependence \cite{Arnhold99,Chicharro09} to phase synchronization \cite{Rosenblum01,Sun12}. Here, we consider the formulation of Granger causality in terms of information flow from $X$ to $Y$, which encompasses linear and nonlinear dynamic systems as well as stochastic processes (for a very recent controversial stand on Granger causality see \cite{Sugihara12}).

The most prominent quantification of information flow from $X$ to $Y$ is by the conditional mutual information (CMI), expressing the amount of information of the future state of $Y$ explained by the current state of $X$ accounting for the current state of $Y$. This was first proposed by Schreiber \cite{Schreiber00} under the name transfer entropy (TE) using standard delay embedding of the current states, and was further modified for non-uniform embedding \cite{Vlachos10,Runge12}. CMI and subsequently mutual information (MI) and entropy bear different expressions for discrete and continuous variables, the former being simpler to estimate. Conversion from continuous to discrete variables by binning techniques is not recommended when the variables are high-dimensional because the estimation of information measures bears the same problems of estimating high dimensional density with a binning method \cite{Scott05}. A different and more efficient discretization can be obtained by using the rank order of the delay components of the high-dimensional variables obtained by delay embedding, an idea first formulated for the entropy, called permutation entropy \cite{Bandt02,Zanin12}. Similar approaches have been developed for TE, such as the symbolic transfer entropy (STE) \cite{Staniek08}, which was further corrected to the so-called transfer entropy on rank vectors (TERV) \cite{Kugiumtzis12}, and the momentary information transfer \cite{Pompe11} (see also \cite{Bahraminasab08} for a combination of permutation entropy and the directional synchronization index).

A first contribution of this work is to extend TERV to be able to quantify the direct information flow from $X$ to $Y$ in the presence of other interacting variables called confounding variables, stacked in the variable $Z$, termed partial TERV (PTERV). The extension is straightforward and is the same as for TE, the partial TE (PTE) \cite{Vakorin09,Papana11b}, and for STE, the partial STE (PSTE) \cite{Papana13}. 

For discrete variables and processes, the equivalence of entropy, in the form of Kolmogorov-Sinai entropy, and  permutation entropy has been established theoretically under different conditions \cite{Amigo05,Haruna11,Keller12,Amigo12}. A recent study carries the equivalence to TE and TERV (as well as STE) but at the asymptotic rate \cite{Haruna12}. These results pave the way to take asymptotic properties of estimates of entropy and MI of discrete variables over to the estimates of entropy and MI of ranked versions of continuous variables, the latter being the points obtained from time series by delay embedding. A second contribution of this work is to derive estimates for the bias and variance of rank-based coupling measures (TERV, STE as well as PTERV and PSTE), form parametric tests of significance of TERV and STE and compare them with randomization tests using time-shifted surrogates, a common approach for nonlinear coupling measures \cite{QuianQuiroga02b}.

Comparative studies have found that TE, and subsequently PTE, perform consistently well, \cite{Lungarella07,Palus07,Papana08b}. Therefore we compare PTERV to PSTE using parametric and randomization significance tests and also to PTE using the randomization test on simulated known coupled chaotic systems. We consider also the presence of stochastic trends in the time series, a situation met often in many applications.

The structure of the paper is as follows. In Section~\ref{sec:Measures}, we present first the measures of information flow for bivariate and multivariate time series and then the partial transfer entropy on rank vectors (PTERV). We present the asymptotic statistical properties of PTERV and the parametric and randomization significance tests for PTERV in Section~\ref{sec:Significance}. Then in Section~\ref{sec:Simulations}, the significance tests for PTERV and PSTE are compared on simulated systems and the two measures are further compared to PTE as well. Discussion of the resutls and concluding remarks are given in Section~\ref{sec:Discussion}.

\section{Transfer Entropy and Rank Vectors}
\label{sec:Measures}

We briefly present first the information measures for bivariate time series $\{x_t,y_t\}_{t=1}^N$, i.e. transfer entropy (TE), symbolic transfer entropy (STE) and transfer entropy on rank vectors (TERV).
Assuming standard delay embedding with the same embedding dimension $m$ and delay $\tau$, a favorable choice
for investigating coupling \cite{Papana11}, the reconstructed points from the two time series are
$\mathbf{x}_t = [x_t,x_{t-\tau},\ldots,x_{t-(m-1)\tau}]$
and $\mathbf{y}_t = [y_t,y_{t-\tau},\ldots,y_{t-(m-1)\tau}]$, respectively. We define the measures for the information flow, or Granger causality, from $X$ to $Y$, denoted $X \rightarrow Y$, assuming the driving system being represented by $X$ and the response system by $Y$. The future state of the response is defined in terms of $T$ times ahead denoted $\mathbf{y}_t^T = [y_{t+1},\ldots,y_{t+T}]$, where often the future horizon is limited to $T=1$ ($\mathbf{y}_t^T = y_{t+1}$).

\subsection{Transfer Entropy}
\label{subsec:TE}

Transfer entropy (TE) is the conditional mutual information $I(\mathbf{y}_t^T ; \mathbf{x}_t | \mathbf{y}_t)$ and quantifies the information about the future of the response system, $\mathbf{y}_t^T$, obtained by the current state of the driving system, $\mathbf{x}_t$, that is not already contained in the current state of the response system, $\mathbf{y}_t$. In terms of entropy, TE is defined as
\begin{equation}
\mbox{TE}_{X\rightarrow Y} = I(\mathbf{y}_t^T ; \mathbf{x}_t | \mathbf{y}_t) =
  -H(\mathbf{y}_t^T,\mathbf{x}_t,\mathbf{y}_t) + H(\mathbf{x}_t,\mathbf{y}_t) + H(\mathbf{y}_t^T,\mathbf{y}_t) -H(\mathbf{y}_t),
\label{eq:TEH}
\end{equation}
where $H(\mathbf{x}) = \int_{\mathbb{X}} f(\mathbf{x}) \ln f(\mathbf{x}) \mbox{d}\mathbf{x}$ is the differential entropy of a continuous variable $\mathbf{x}$ with domain $\mathbb{X}$, and $f(\mathbf{x})$ is the probability density function of $\mathbf{x}$. In estimating $\mbox{TE}_{X\rightarrow Y}$ one can assume discretization of the observed variables $x_t$ and $y_t$ and use the Shannon entropy $H(\mathbf{x}) = \sum p(\mathbf{x})\ln p(\mathbf{x})$ for the discrete variable $\mathbf{x}$, where the sum is over the possible bins of $\mathbf{x}$ and $p(\mathbf{x})$ is the probability mass function (pmf) of $\mathbf{x}$. However, binning methods are found to be more demanding on data size than other methods approximating directly the density function, and subsequently the differential entropy. In particular, for high dimensions, i.e. large $m$, the $k$-nearest neighbor estimate turns out to be the most robust to time series length \cite{Kraskov04,Vlachos10}, and we apply this in the comparative study in Section~\ref{sec:Simulations}.

The inefficiency of the binning methods for estimating entropies is attributed to the bias due to the estimation of bin probability with the relative frequency of occurrence of entries in the bin, and the variance due to having a number of unpopulated or poorly populated bins. The latter increases with the embedding dimension $m$, and it is noted that for the discretization of $x_t$ and $y_t$ in $b$ bins the variable of highest dimension $[\mathbf{y}_t^T,\mathbf{x}_t,\mathbf{y}_t]$ in eq.(\ref{eq:TEH}) regards $b^{2m+T}$ bins.

\subsection{Symbolic Transfer Entropy}
\label{subsec:STE}
A different discretization that produces far less bins for the high dimensional variables is provided by the rank ordering of the components of vector variables. For each point $\mathbf{y}_t$, the ranks of its components, say in ascending order, form a rank vector $\hat{\mathbf{y}}_t = [r_{t,1},r_{t,2},\ldots,r_{t,m}]$, where $r_{t,j}\in\{1,2,\ldots,m\}$ for $j=1,\ldots,m$, is the rank order of the component $y_{t-(j-1)\tau}$. For two equal components of $\mathbf{y}_t$ the smallest rank is assigned to the component appearing first in $\mathbf{y}_t$. Substituting rank vectors to sample vectors in the expression for Shannon entropy gives the so-called permutation entropy $H(\hat{\mathbf{x}}) = \sum p(\hat{\mathbf{x}})\ln p(\hat{\mathbf{x}})$, where the sum is over $m!$ possible permutations of the $m$ components of $\hat{\mathbf{x}}$ \cite{Bandt02}.

The same conversion has been suggested for TE.
In \cite{Staniek08}, the arguments in the CMI of $\mbox{TE}_{X\rightarrow Y}$ in eq.(\ref{eq:TEH}) are modified as follows: $\mathbf{x}_t$ and $\mathbf{y}_t$ are substituted by the respective rank vectors $\hat{\mathbf{x}}_t$ and $\hat{\mathbf{y}}_t$, and the future response vector $\mathbf{y}_t^T$ is replaced by the response rank vector at time $t+T$, $\hat{\mathbf{y}}_{t+T}$. This conversion of TE is called symbolic transfer entropy (STE) defined as
\begin{equation}
\mbox{STE}_{X\rightarrow Y} = I(\hat{\mathbf{y}}_{t+T} ; \hat{\mathbf{x}}_t | \hat{\mathbf{y}}_t) =  -H(\hat{\mathbf{y}}_{t+T},\hat{\mathbf{x}}_t,\hat{\mathbf{y}}_t) + H(\hat{\mathbf{x}}_t,\hat{\mathbf{y}}_t) + H(\hat{\mathbf{y}}_{t+T},\hat{\mathbf{y}}_t) -H(\hat{\mathbf{y}}_t).
\label{eq:STEH}
\end{equation}

\subsection{Transfer Entropy on Rank Vectors}
\label{subsec:TERV}
STE is not the direct analogue to TE in terms of ranks. While the correspondence of $\mathbf{y}_t$ to $\hat{\mathbf{y}}_t$ (and the same for $\mathbf{x}_t$ and $\hat{\mathbf{x}}_t$) is direct and preserves the vector dimension, the vector $\mathbf{y}_t^T = [y_{t+1},\ldots,y_{t+T}]$ of dimension $T$ is mapped to $\hat{\mathbf{y}}_{t+T} = [r_{t+T,1},\ldots,r_{t+T,m}]$ of dimension $m$, the rank vector of $\mathbf{y}_{t+T} = [y_{t+T},\ldots,y_{t+T-(m-1)\tau}]$. This indirect correspondence has implications in the computation of the entropy terms and thus the CMI. In particular, supposing $\tau=1$, the joint vector $[\mathbf{y}_{t+T}, \mathbf{y}_t]$ present in two entropy terms of eq.(\ref{eq:TEH}) has $m+T$ consecutive components, and the possible rank orders of the corresponding rank vector are $(m+T)!$. Using the approach in STE, $[\mathbf{y}_{t+T}, \mathbf{y}_t]$ is mapped to $[\hat{\mathbf{y}}_{t+T}, \hat{\mathbf{y}}_t]$ having $m!\cdot\frac{m!}{(m-T)!}$ possible rank orders (for details see \cite{Kugiumtzis12}).

In \cite{Kugiumtzis12}, a correction of STE termed transfer entropy on rank vectors (TERV) is proposed. TERV assigns for the future response sample vector $\mathbf{y}_t^T=[y_{t+1},\ldots,y_{t+T}]$ in TE the future response rank vector $\hat{\mathbf{y}}^T_{t}=[r_{t,m+1},\ldots,r_{t,m+T}]$ containing the ranks of $[y_{t+1},\ldots,y_{t+T}]$ in the augmented vector $[\mathbf{y}_t,\mathbf{y}_t^T]$. TERV is thus defined as
\begin{equation}
\mbox{TERV}_{X\rightarrow Y} = I(\hat{\mathbf{y}}^T_{t} ; \hat{\mathbf{x}}_t | \hat{\mathbf{y}}_t) = -H(\hat{\mathbf{y}}^T_{t},\hat{\mathbf{x}}_t,\hat{\mathbf{y}}_t) + H(\hat{\mathbf{x}}_t,\hat{\mathbf{y}}_t) + H(\hat{\mathbf{y}}^T_{t},\hat{\mathbf{y}}_t) -H(\hat{\mathbf{y}}_t). \nonumber
\label{eq:TERVH}
\end{equation}
We note that there is one-to-one mapping between the ranks of $\mathbf{y}_t$ and the ranks of the first $m$ components of the augmented vector $[\mathbf{y}_t,\mathbf{y}_t^T]$ \footnote{For both there is a bijection to the same rank sequence as defined in \cite{Amigo05,Haruna12}.}. Thus the pmf for the first $m$ components of the rank vector of $[\mathbf{y}_t,\mathbf{y}_t^T]$ is the same as the pmf for the ranks of $\mathbf{y}_t$, defined on $m!$ possible rank vectors. For the rest $T$ components of the rank vector of $[\mathbf{y}_t,\mathbf{y}_t^T]$, there are $\prod_{i=1}^T(m+i)$ possible combinations, giving a total number of $(m+T)!$ combinations for the rank vector of $[\mathbf{y}_t,\mathbf{y}_t^T]$. This is larger than the number of combinations STE assigns for $[\mathbf{y}_t,\mathbf{y}_t^T]$, $(m+T)! > m!\cdot\frac{m!}{(m-T)!}$, and thus the entropy terms $H(\mathbf{y}_t^T,\mathbf{x}_t,\mathbf{y}_t)$ and $H(\mathbf{y}_t^T,\mathbf{y}_t)$ for the continuous variables in eq.(\ref{eq:TEH}) for TE are correctly estimated by TERV but underestimated by STE, having $H(\hat{\mathbf{y}}_{t+T},\hat{\mathbf{x}}_t,\hat{\mathbf{y}}_t) < H(\hat{\mathbf{y}}^T_{t},\hat{\mathbf{x}}_t,\hat{\mathbf{y}}_t)$ and $H(\hat{\mathbf{y}}_{t+T},\hat{\mathbf{y}}_t) < H(\hat{\mathbf{y}}^T_{t},\hat{\mathbf{y}}_t)$  \cite{Kugiumtzis12}.

\subsection{Partial Transfer Entropy on Rank Vectors (PTERV)}
\label{subsec:PTERV}

Now we suppose we have $K$ multivariate time series, denoted $\{x_t,y_t,z_{1,t},\ldots,z_{K-2,t}\}_{t=1}^N$, in order to preserve the notation for the relation of interest $X \rightarrow Y$ in the presence of $K-2$ other observed variables $Z_1,\ldots,Z_{K-2}$. For direct Granger causality or direct information flow from $X$ to $Y$, one has to account for the confounding variables $Z=\{Z_1,\ldots,Z_{K-2}\}$. The delay embedding for each $Z_i$, $i=1,\ldots,K-2$, gives $\mathbf{z}_{i,t} = [z_{i,t},z_{i,t-\tau},\ldots,z_{i,t-(m-1)\tau}]$, and for convenience we stack all the reconstructed points $\mathbf{z}_{i,t}$ to one $\mathbf{z}_{t}=[\mathbf{z}_{1,t},\ldots,\mathbf{z}_{K-2,t}]$ of dimension $(K-2)m$. 

TE has been extended to include the effect of the current state of $Z$ on the future of the response $Y$ and the current state of $X$, simply by adding it to the conditioning term of CMI. The so-called partial transfer entropy (PTE) is defined as \cite{Vakorin09,Papana11b}
\begin{equation}
 \begin{array}{rcl}
 \mbox{PTE}_{X\rightarrow Y | Z} & = & I(\mathbf{y}_t^T ; \mathbf{x}_t | \mathbf{y}_t,\mathbf{z}_{t}) \\
  & = & -H(\mathbf{y}_t^T,\mathbf{x}_t,\mathbf{y}_t,\mathbf{z}_{t}) + H(\mathbf{x}_t,\mathbf{y}_t,\mathbf{z}_{t}) + H(\mathbf{y}_t^T,\mathbf{y}_t,\mathbf{z}_{t}) -H(\mathbf{y}_t,\mathbf{z}_{t}).
 \end{array}
\label{eq:PTEH}
\end{equation}
The same formulation can be considered for STE and the partial symbolic transfer entropy (PSTE) is defined as \cite{Papana13}
\begin{equation}
 \begin{array}{rcl}
\mbox{PSTE}_{X\rightarrow Y | Z} & = & I(\hat{\mathbf{y}}_{t+T} ; \hat{\mathbf{x}}_t | \hat{\mathbf{y}}_t, \hat{\mathbf{z}}_t) \\
 & = & -H(\hat{\mathbf{y}}_{t+T},\hat{\mathbf{x}}_t,\hat{\mathbf{y}}_t,\hat{\mathbf{z}}_t) + H(\hat{\mathbf{x}}_t,\hat{\mathbf{y}}_t,\hat{\mathbf{z}}_t) + H(\hat{\mathbf{y}}_{t+T},\hat{\mathbf{y}}_t,\hat{\mathbf{z}}_t) -H(\hat{\mathbf{y}}_t,\hat{\mathbf{z}}_t).
 \end{array}
\label{eq:PSTEH}
\end{equation}
Here we extend also TERV to the partial transfer entropy on rank vectors (PTERV) defined analogously as
\begin{equation}
 \begin{array}{rcl}
    \mbox{PTERV}_{X\rightarrow Y | Z} & = & I(\hat{\mathbf{y}}^T_{t} ; \hat{\mathbf{x}}_t | \hat{\mathbf{y}}_t,\hat{\mathbf{z}}_t) \\
  & = & -H(\hat{\mathbf{y}}^T_{t},\hat{\mathbf{x}}_t,\hat{\mathbf{y}}_t,\hat{\mathbf{z}}_t) + H(\hat{\mathbf{x}}_t,\hat{\mathbf{y}}_t,\hat{\mathbf{z}}_t) + H(\hat{\mathbf{y}}^T_{t},\hat{\mathbf{y}}_t,\hat{\mathbf{z}}_t) -H(\hat{\mathbf{y}}_t,\hat{\mathbf{z}}_t).
 \end{array}
\label{eq:PTERVH}
\end{equation}
The dimension of the variables in the entropy terms for all expressions (PTE, PSTE and PTERV) increase with the number of variables $K$ and the embedding dimension $m$. For PSTE and PTERV the possible rank permutations of $\mathbf{z}_{t}=[\mathbf{z}_{1,t},\ldots,\mathbf{z}_{K-2,t}]$ are $(m!)^{K-2}$, which shows that the demand for data size increases sharply with both $K$ and $m$.

\section{Statistical Significance of PTERV}
\label{sec:Significance}

A main disadvantage of nonlinear measures of Granger causality as opposed to linear causality measures is the lack of established asymptotic properties. However, dealing with discretized variables, as for TERV and PTERV (and respectively STE and PSTE), it is possible to do so, based on the estimation of Shannon entropy. There have been a number of works on estimates of Shannon entropy and their statistical properties, and subsequently for mutual information \cite{Miller55,Grassberger88,Pardo95,Roulston99,Antos01,Grassberger03,Paninski03,Schuermann04,Goebel05,Hutter05,Bonachela08,Lesne09,Vinck12}.
\subsection{Bias and variance in the estimation of entropy}
\label{subsec:Hbiasvar}

Here we follow the approximation for the bias and variance of Shannon entropy in \cite{Miller55} and \cite{Roulston99}, and further extend it including higher order terms of the Taylor series expansion.

Let us suppose that for the discrete variable $X$ there are $B$ states. For our setting $X$ can be, say, $\mathbf{x}_t$, and then $B=m!$. Let the true probability for each state $i=1,\ldots,B$, be $p_i$, the observed frequency of state $i$ be $n_i$, and the estimated probability of state $i$ be $q_i = n_i/N$, where $N$ is the data size. The relative error in the estimation of the probability of state $i$ is $\epsilon_i=(q_i - p_i)/p_i$.
The following approach is based on the assumption that $n_i$ is a binomial random variable, $n_i \sim \mathbb{B}(N,p_i)$.

The observed Shannon entropy is
\begin{eqnarray*}
    H_{\ssmall{obs}} & = & - \sum_{i=1}^B q_i \ln q_i = - \sum_{i=1}^B p_i(1+\epsilon_i) \ln(p_i(1+\epsilon_i))  \\
    & = & - \sum_{i=1}^B p_i \ln p_i - \sum_{i=1}^B \left( p_i\epsilon_i\ln p_i + p_i(1+\epsilon_i) \ln (1+\epsilon_i) \right),
\end{eqnarray*}
where the first term in the last expression is the true entropy $H_{\infty}$. Using the Taylor expansion up to third order of $\ln(1+\epsilon_i)$, and taking expectation we arrive at the expression for the bias
\begin{equation}
\mbox{B}(H_{\ssmall{obs}}) = \langle H_{\ssmall{obs}} \rangle - H_{\infty} = - \sum_{i=1}^B \left( \frac{p_i}{2} \langle \epsilon_i^2 \rangle - \frac{p_i}{6} \langle \epsilon_i^3 \rangle \right).
\label{eq:Hobs2}
\end{equation}
Making use of the first three moments of the binomial distribution, we have $\langle \epsilon_i^2 \rangle = (1-p_i)/(Np_i)$ and $\langle \epsilon_i^3 \rangle = (1-3p_i+2p_i^2)/(N^2p_i^2)$, and substituting them in eq.(\ref{eq:Hobs2}) we get
\begin{equation}
\mbox{B}(H_{\ssmall{obs}}) = - \frac{B^*-1}{2N} - \frac{3B^*-2}{6N^2} + \frac{1}{6N} \sum_{i=1}^{B^*}\frac{1}{n_i},
\label{eq:Hbias}
\end{equation}
where $B^*$ is the number of states with positive observed frequency. The first term in the rhs of eq.(\ref{eq:Hbias}) is the expression added to $H_{\ssmall{obs}}$ to give the bias-corrected entropy estimate of Miller-Madow \cite{Miller55}, which we expand here to third order Taylor approximation. The information from the observed variable enters in the bias approximation of Miller-Madow only in terms of the number of observed states $B^*$, but in eq.(\ref{eq:Hbias}) also the observed frequencies are considered. However, the effect of the two additional terms is downweighted by the data size $N$, and the bias in eq.(\ref{eq:Hbias}) converges to $-(B^*-1)/(2N)$ with the increase of $N$.

For the variance of the Shannon entropy estimate, we adopt the approximation given in \cite{Roulston99}, where the error propagation formula
\[
\mbox{Var}(H_{\ssmall{obs}}) = \sum_{i=1}^{B} \left(\frac{\partial H_{\ssmall{obs}}}{\partial n_i} \right)^2 \mbox{Var}(n_i)
\]
is invoked to arrive at the expression
\begin{equation}
\mbox{Var}(H_{\ssmall{obs}}) = \frac{1}{N}\sum_{i=1}^{B} (\ln q_i + H_{\ssmall{obs}})^2q_i(1-q_i).
\label{eq:Hvariance}
\end{equation}
We note that for both bias and variance formulas in eq.(\ref{eq:Hbias}) and eq.(\ref{eq:Hvariance}), respectively, the independence of the binomial variables $n_i$ is assumed.

\subsection{Bias and variance of PTERV}
\label{subsec:PTERVbiasvar}

In order to derive the bias for PTERV, we first recall that the number of possible states for $\hat{\mathbf{y}}^T_{t}$ is $B_1 = \prod_{i=1}^T(m+i)$, for $\hat{\mathbf{x}}_t$ is $B_2 = m!$, for $\hat{\mathbf{y}}_t$ is $B_3=m!$, and for $\hat{\mathbf{z}}_t$ is $B_4=(m!)^{K-2}$. These variables enter jointly in the entropy terms of the expression of PTERV in eq.(\ref{eq:PTERVH}), and the possible states of the joint vector variables are as follows: $B_{1234}=B_1 B_2 B_3 B_4$ for $[\hat{\mathbf{y}}^T_{t}, \hat{\mathbf{x}}_t, \hat{\mathbf{y}}_t, \hat{\mathbf{z}}_t]$, $B_{234}= B_2 B_3 B_4$ for $[\hat{\mathbf{x}}_t, \hat{\mathbf{y}}_t, \hat{\mathbf{z}}_t]$, $B_{134}=B_1 B_3 B_4$ for $[\hat{\mathbf{y}}^T_{t}, \hat{\mathbf{y}}_t, \hat{\mathbf{z}}_t]$, and $B_{34}= B_3 B_4$ for $[\hat{\mathbf{y}}_t, \hat{\mathbf{z}}_t]$. The observed frequency and the estimated probability for $[\hat{\mathbf{y}}^T_{t}, \hat{\mathbf{x}}_t, \hat{\mathbf{y}}_t, \hat{\mathbf{z}}_t]$ are denoted as $n_{ijkl}$ and $q_{ijkl}=n_{ijkl}/(N-m-T)$, respectively, where the index $i=1,\ldots,B_1$, denotes the state of $\hat{\mathbf{y}}^T_{t}$, and respectively for the other indices. The observed frequency and estimated probability for the other three variables of smaller dimension are denoted in terms of the respective marginal distributions, e.g. $n_{..kl}$ and $q_{..kl}$ for $[\hat{\mathbf{y}}_t, \hat{\mathbf{z}}_t]$. For each vector variable, the number of states with positive observed frequency is denoted with an asterisk in superscript, e.g. $B_{34}^*$ for the active states of $[\hat{\mathbf{y}}_t, \hat{\mathbf{z}}_t]$.

Substituting the bias expression in eq.(\ref{eq:Hbias}) for each entropy term in the expression for PTERV in eq.(\ref{eq:PTERVH}), we get the expression for the bias of PTERV
\begin{eqnarray}
  & & \mbox{B}(I_{\ssmall{obs}}(\hat{\mathbf{y}}^T_{t} ; \hat{\mathbf{x}}_t | \hat{\mathbf{y}}_t,\hat{\mathbf{z}}_t)) = \langle I_{\ssmall{obs}}(\hat{\mathbf{y}}^T_{t} ; \hat{\mathbf{x}}_t | \hat{\mathbf{y}}_t,\hat{\mathbf{z}}_t)  \rangle -
     I_{\infty}(\hat{\mathbf{y}}^T_{t} ; \hat{\mathbf{x}}_t | \hat{\mathbf{y}}_t,\hat{\mathbf{z}}_t) =
     \nonumber \\
    & & \frac{B_{1234}^*-B_{234}^*-B_{134}^*+B_{34}^*-4}{2(N-m-T)} + \frac{3B_{1234}^*-3B_{234}^*-3B_{134}^*+3B_{34}^*-8}{6(N-m-T)^2}
\label{eq:CMIbias} \\
    & & + \frac{1}{6(N-m-T)} \sum_{k=1}^{B_3^*} \sum_{l=1}^{B_4^*} \left(-\sum_{i=1}^{B_1^*} \sum_{j=1}^{B_2^*}\frac{1}{n_{ijkl}} + \sum_{j=1}^{B_2^*}\frac{1}{n_{.jkl}} +  \sum_{i=1}^{B_1^*}\frac{1}{n_{i.kl}} - \frac{1}{n_{..kl}} \right).
\nonumber
\end{eqnarray}

For the variance of PTERV, we use the error propagation formula for $I_{\ssmall{obs}}(\hat{\mathbf{y}}^T_{t} ; \hat{\mathbf{x}}_t | \hat{\mathbf{y}}_t,\hat{\mathbf{z}}_t)$
\[
\mbox{Var}(I_{\ssmall{obs}}(\hat{\mathbf{y}}^T_{t} ; \hat{\mathbf{x}}_t | \hat{\mathbf{y}}_t,\hat{\mathbf{z}}_t)) = \sum_{i=1}^{B_1^*} \sum_{j=1}^{B_2^*} \sum_{k=1}^{B_3^*} \sum_{l=1}^{B_4^*} \left(\frac{\partial I_{\ssmall{obs}}(\hat{\mathbf{y}}^T_{t} ; \hat{\mathbf{x}}_t | \hat{\mathbf{y}}_t,\hat{\mathbf{z}}_t)}{\partial n_{ijkl}} \right)^2 \mbox{Var}(n_{ijkl}).
\]
Following the same approximation as for the mutual information of scalar variables in \cite{Roulston99}, we arrive at the expression for the variance
\begin{eqnarray}
& & \mbox{Var}(I_{\ssmall{obs}}(\hat{\mathbf{y}}^T_{t} ; \hat{\mathbf{x}}_t | \hat{\mathbf{y}}_t,\hat{\mathbf{z}}_t)) = \frac{1}{N-m-T}\sum_{i=1}^{B_1^*} \sum_{j=1}^{B_2^*} \sum_{k=1}^{B_3^*} \sum_{l=1}^{B_4^*} \left( -\ln q_{ijkl} \right.
     \nonumber \\
& & \quad + \left. \ln q_{.jkl} + \ln q_{i.kl} - \ln q_{..kl} +  I_{\ssmall{obs}}(\hat{\mathbf{y}}^T_{t} ; \hat{\mathbf{x}}_t | \hat{\mathbf{y}}_t,\hat{\mathbf{z}}_t) \right)^2 q_{ijkl}(1-q_{ijkl}).
\label{eq:CMIvariance}
\end{eqnarray}

\subsection{Parametric Significance Test for PTERV}
\label{subsec:partest}

We consider now parametric tests for the significance of PTERV, i.e. for the null hypothesis H$_0$: $I_{\infty}(\hat{\mathbf{y}}^T_{t} ; \hat{\mathbf{x}}_t | \hat{\mathbf{y}}_t,\hat{\mathbf{z}}_t)=0$, indicating that the variables $\hat{\mathbf{y}}^T_{t}$ and $\hat{\mathbf{x}}_t$ are independent conditioned on the variable $(\hat{\mathbf{y}}_t,\hat{\mathbf{z}}_t)$. A natural choice for the asymptotic distribution of the estimate of mutual information is Gaussian, e.g. see \cite{Paninski03,Hutter05},
and employing the results for the bias in eq.(\ref{eq:CMIbias}) and variance in eq.(\ref{eq:CMIvariance}), we have the first candidate for the null distribution of the PTERV statistic
\begin{equation}
I_{\ssmall{obs}}(\hat{\mathbf{y}}^T_{t} ; \hat{\mathbf{x}}_t | \hat{\mathbf{y}}_t,\hat{\mathbf{z}}_t) \sim
\mbox{N}\left(
 \mbox{B}(I_{\ssmall{obs}}(\hat{\mathbf{y}}^T_{t} ; \hat{\mathbf{x}}_t | \hat{\mathbf{y}}_t,\hat{\mathbf{z}}_t)), \mbox{Var}(I_{\ssmall{obs}}(\hat{\mathbf{y}}^T_{t} ; \hat{\mathbf{x}}_t | \hat{\mathbf{y}}_t,\hat{\mathbf{z}}_t)) \right).
\label{eq:partestnormal}
\end{equation}

The mutual information of two independent discrete variables $X$ and $Y$ is related to the Chi-square statistic for the independence test with the expression $\chi^2 = 2 N I_{\ssmall{obs}}(X;Y)$, where the Chi-square distribution has $(|X|-1)(|Y|-1)$ degrees of freedom (dof), $|X|$ is the number of states of $X$ and $N$ is the number of observations of $X$ and $Y$ \cite{Pardo95,Goebel05}. Based on this result, Goebel et al \cite{Goebel05} proved that the statistic $I_{\ssmall{obs}}(X;Y)$ follows Gamma distribution. They extended their result to the conditional mutual information, and proved that $I(X;Y|Z)$ when $X$ and $Y$ are independent conditioned on $Z$ follows also Gamma distribution with shape parameter $\kappa=|Z|(|X|-1)(|Y|-1)/2$ and scale parameter $\theta=1/N$. This result adapted to our setting constitutes the second candidate for the null distribution of PTERV
\begin{equation}
I_{\ssmall{obs}}(\hat{\mathbf{y}}^T_{t} ; \hat{\mathbf{x}}_t | \hat{\mathbf{y}}_t,\hat{\mathbf{z}}_t) \sim
\Gamma\left(\frac{B^*_{34}}{2}(B^*_1-1)(B^*_2-1), \frac{1}{(N-m-T)} \right).
\label{eq:partestgamma1}
\end{equation}

If instead we trust the bias and variance approximation in eq.(\ref{eq:CMIbias}) and eq.(\ref{eq:CMIvariance}), we can derive the shape and scale parameter of the Gamma null distribution from these, and then we have the third candidate for the null distribution of PTERV
\begin{equation}
I_{\ssmall{obs}}(\hat{\mathbf{y}}^T_{t} ; \hat{\mathbf{x}}_t | \hat{\mathbf{y}}_t,\hat{\mathbf{z}}_t) \sim
\Gamma\left(\frac{\mbox{B}(I_{\ssmall{obs}}(\hat{\mathbf{y}}^T_{t} ; \hat{\mathbf{x}}_t | \hat{\mathbf{y}}_t,\hat{\mathbf{z}}_t))^2}{\mbox{Var}(I_{\ssmall{obs}}(\hat{\mathbf{y}}^T_{t} ; \hat{\mathbf{x}}_t | \hat{\mathbf{y}}_t,\hat{\mathbf{z}}_t)}, \frac{\mbox{Var}(I_{\ssmall{obs}}(\hat{\mathbf{y}}^T_{t} ; \hat{\mathbf{x}}_t | \hat{\mathbf{y}}_t,\hat{\mathbf{z}}_t))}{\mbox{B}(I_{\ssmall{obs}}(\hat{\mathbf{y}}^T_{t} ; \hat{\mathbf{x}}_t | \hat{\mathbf{y}}_t,\hat{\mathbf{z}}_t))} \right).
\label{eq:partestgamma2}
\end{equation}
We note that the Gamma approximation of the null distribution of PTERV in eq.(\ref{eq:partestgamma2}), termed Gamma-2, converges to the Gaussian distribution in eq.(\ref{eq:partestnormal}) with the increase of bias, while the Gamma approximation in eq.(\ref{eq:partestgamma1}), termed Gamma-1, can differ in mean from Gamma-2 and Gaussian distributions.

\begin{figure}
\centering
\hbox{\includegraphics[width=6cm]{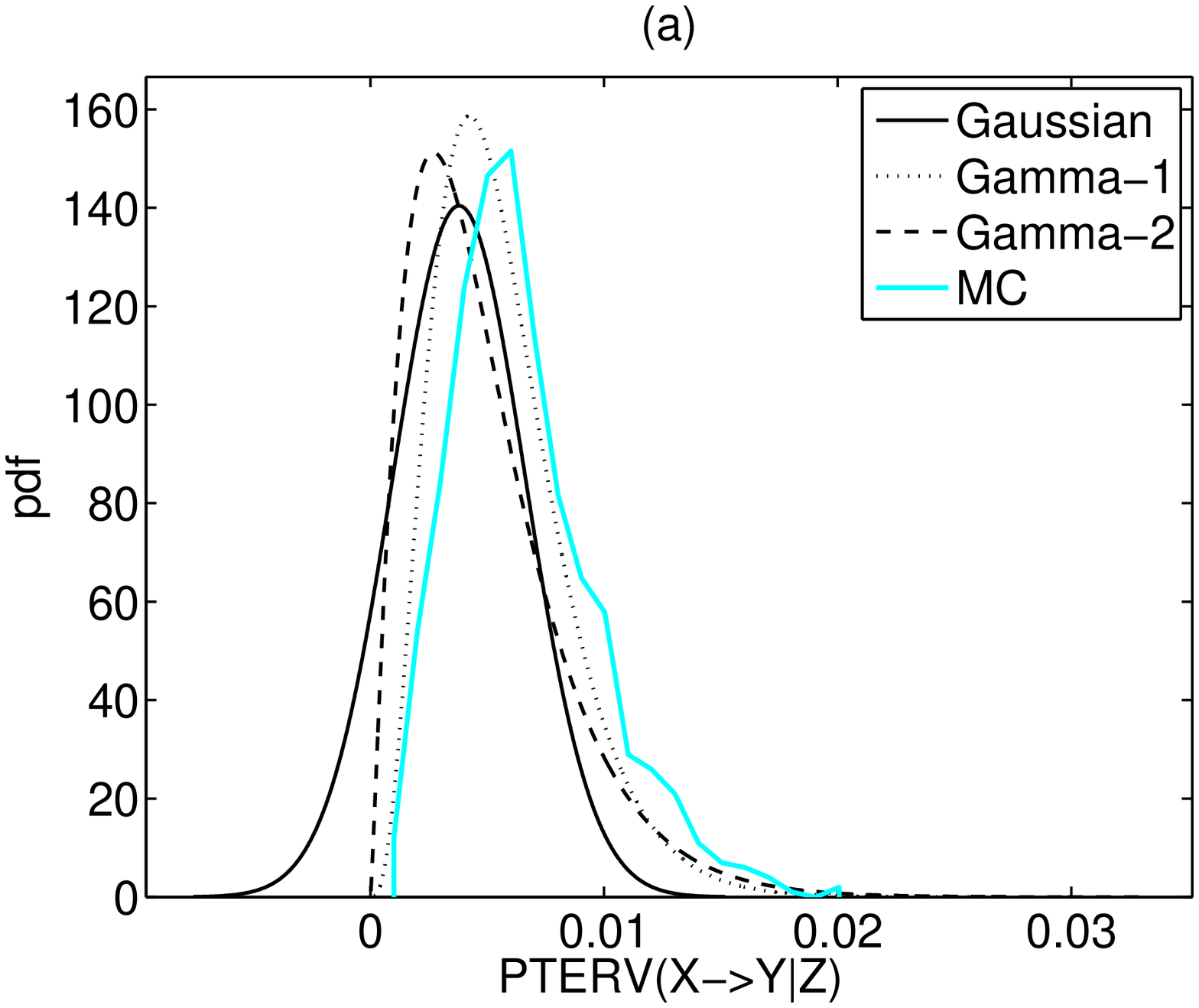}
\includegraphics[width=6cm]{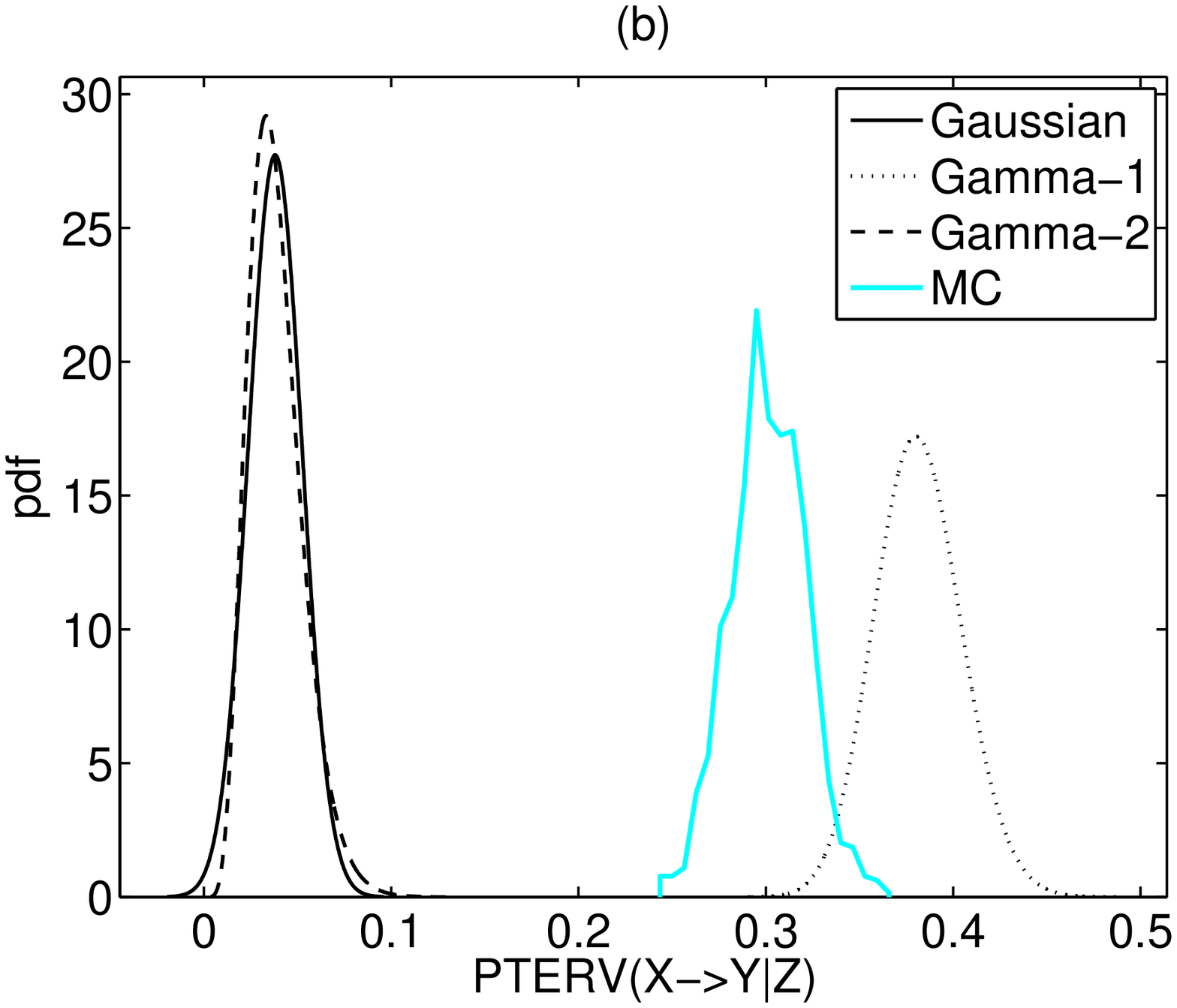}}
\caption{The true distribution of PTERV and the three
approximations, as given in the legend, formed from 1000 Monte Carlo realizations of the exemplary system of two
independent variables conditioned on a third variable. The parameters are $T=1$, $N=1024$, $m=2$ in (a) and $m=3$ in (b).}
\label{fig:pardistrPTERV}       
\end{figure}

We show the differences of Gaussian, Gamma-1 and Gamma-2
distributions in approximating the true PTERV distribution with a
simple example representing the null hypothesis of conditional
independence. We generate the conditioning variable $Z$ from a
first order autoregressive process and let the other two variables
depend linearly on $Z$
\begin{eqnarray}
x_t & = & a z_t + \epsilon_{x,t} \nonumber \\
y_t & = & b z_t + c x_{t-1} + \epsilon_{y,t}
\label{eq:example1} \\
z_t & = & d z_{t-1} + \epsilon_{z,t} \nonumber
\end{eqnarray}
where $a=2$, $b=-1$, $d=0.8$, and $\epsilon_{x,t}$,
$\epsilon_{y,t}$, $\epsilon_{z,t}$ are Gaussian white noise with
standard deviation (SD) 1, 2 and 1, respectively. We set $c=0$ to
have $X$ and $Y$ conditionally independent. In
Fig.~\ref{fig:pardistrPTERV}, we show the approximate true
distribution of PTERV ($T=1$) from 1000 Monte Carlo realizations.
We compute the average parameter values from the 1000
realizations for the three distributions in eq.(\ref{eq:partestnormal}), eq.(\ref{eq:partestgamma1}) and eq.(\ref{eq:partestgamma2}), and draw the three approximating distributions.
Gamma-1 matches best the true distribution of PTERV for $m=2$, but
for $m=3$ it lies at the right of the true distribution. For
$m=3$, Gaussian and Gamma-2 distributions converge but lie both far to
the left of the true distribution. The deviation of all
three approximations from the true distribution is larger when we
further increase $m$ or employ a more involved setting, e.g. using
nonlinear dependence of $X$ and $Y$ on $Z$.

First, we attempt to explain the shortcoming of Gamma-1
approximation for increasing $m$. Considering the Chi-square independence test in the
setting of joint rank variables, we expect to have zero cells and
incomplete contingency table, so that the degrees of freedom
$B^*_{34}(B^*_1-1)(B^*_2-1)$ of the  Chi-square distribution
associated with the Gamma-1 distribution in
eq.(\ref{eq:partestgamma1}) may be overestimated. In turn, this
leads to overestimation of the bias
$\mbox{B}(I_{\ssmall{obs}}(\hat{\mathbf{y}}^T_{t} ;
\hat{\mathbf{x}}_t | \hat{\mathbf{y}}_t,\hat{\mathbf{z}}_t))$,
because the mean of the null Gamma distribution is the
product of the shape and scale parameter. To the best of our
knowledge there is no standard approach to determine the degrees
of freedom for the Chi-square statistic for incomplete contingency
tables, especially when the zero cells depend on the correlation
structure of the involved variables, e.g. see \cite{Kim97,Kang12}.

Regarding the Gaussian and Gamma-2 approximations, we first note
that the shape of the approximating distribution to the true null
distribution of PTERV does not constitute an issue. The bias of
PTERV is large (compared to the width of the distribution) even for
small $m$ and $K$ and the distribution of PTERV becomes
essentially Gaussian \cite{Paninski03}. The estimation of the bias
of PTERV, even when extending the Taylor approximation to third
order terms (see eq.(\ref{eq:CMIbias})), is not accurate and
deviates at an extent that depends on the parameters involved in
the definition of PTERV ($m$, $K$ and $T$), and also on the
inter-dependence of the components in
$(\hat{\mathbf{y}}_t,\hat{\mathbf{z}}_t)$ and the dependence of
$\hat{\mathbf{y}}^T_{t}$ and $\hat{\mathbf{x}}_t$ on
$(\hat{\mathbf{y}}_t,\hat{\mathbf{z}}_t)$. These dependencies
cause violation of the assumption of independence of the binomial
variables $n_{ijkl}$ in the computation of the bias and the effect
is larger when the number of states of the involved variables
increase, i.e. either of $m$, $K$ and $T$ increases.

\subsection{Randomization Significance Test for PTERV}
\label{subsec:randtest}

The insufficiency of the analytic approximation of the true null
distribution for the significance test for PTERV paves the way for
employing resampling approaches. We consider here the
randomization test making use of the time-shifted surrogates
\cite{QuianQuiroga02b}. For PTERV, this is succeeded by circular
shifting the components of the time series of the driving
variable, $\{\hat{\mathbf{x}}_t\} =
\{\hat{\mathbf{x}}_{(m-1)\tau+1},\hat{\mathbf{x}}_{(m-1)\tau+2},\ldots,\hat{\mathbf{x}}_{N-T}\}$,
by a random time step $w$ producing the surrogate time series
\[
\{\hat{\mathbf{x}}^{*}_t\} = \{\hat{\mathbf{x}}_{(m-1)\tau+w+1},\ldots,\hat{\mathbf{x}}_{N-T}, \hat{\mathbf{x}}_{(m-1)\tau+1},\ldots,\hat{\mathbf{x}}_{(m-1)\tau+w}\}.
\]
The tuple $\{\hat{\mathbf{y}}^T_{t}, \hat{\mathbf{x}}^*_t,
(\hat{\mathbf{y}}_t,\hat{\mathbf{z}}_t)\}$ represents H$_0$ as the
intrinsic dynamics of each variable are preserved but the coupling
between the driving variable $\hat{\mathbf{x}}^*_t$ and the
response variable $\hat{\mathbf{y}}^T_{t}$ is destroyed, so that
it holds $I_{\infty}(\hat{\mathbf{y}}^T_{t} ; \hat{\mathbf{x}}^*_t
| \hat{\mathbf{y}}_t,\hat{\mathbf{z}}_t) = 0$. However, the
time displacement of $\hat{\mathbf{x}}_t$ destroys also the possible
dependence structure of $\hat{\mathbf{x}}_t$ on the conditioning
variable $(\hat{\mathbf{y}}_t,\hat{\mathbf{z}}_t)$. This is a
point of inconsistency that may affect the matching of the sample
distribution of PTERV, computed on an ensemble of surrogate tuples $\{\hat{\mathbf{y}}^T_{t}, \hat{\mathbf{x}}^*_t,
(\hat{\mathbf{y}}_t,\hat{\mathbf{z}}_t)\}$, to the true null
distribution of $I_{\infty}(\hat{\mathbf{y}}^T_{t} ;
\hat{\mathbf{x}}_t | \hat{\mathbf{y}}_t,\hat{\mathbf{z}}_t)$. On
the other hand, the construction of surrogate tuples
$\{\hat{\mathbf{y}}^T_{t}, \hat{\mathbf{x}}^*_t,
(\hat{\mathbf{y}}_t,\hat{\mathbf{z}}_t)\}$ that preserve all the
original inter-dependencies and destroys only the coupling of
$\hat{\mathbf{y}}^T_{t}$ and $\hat{\mathbf{x}}^*_t$ is an open,
yet unsolved, problem.

\begin{figure}
\centering
\hbox{\includegraphics[width=6cm]{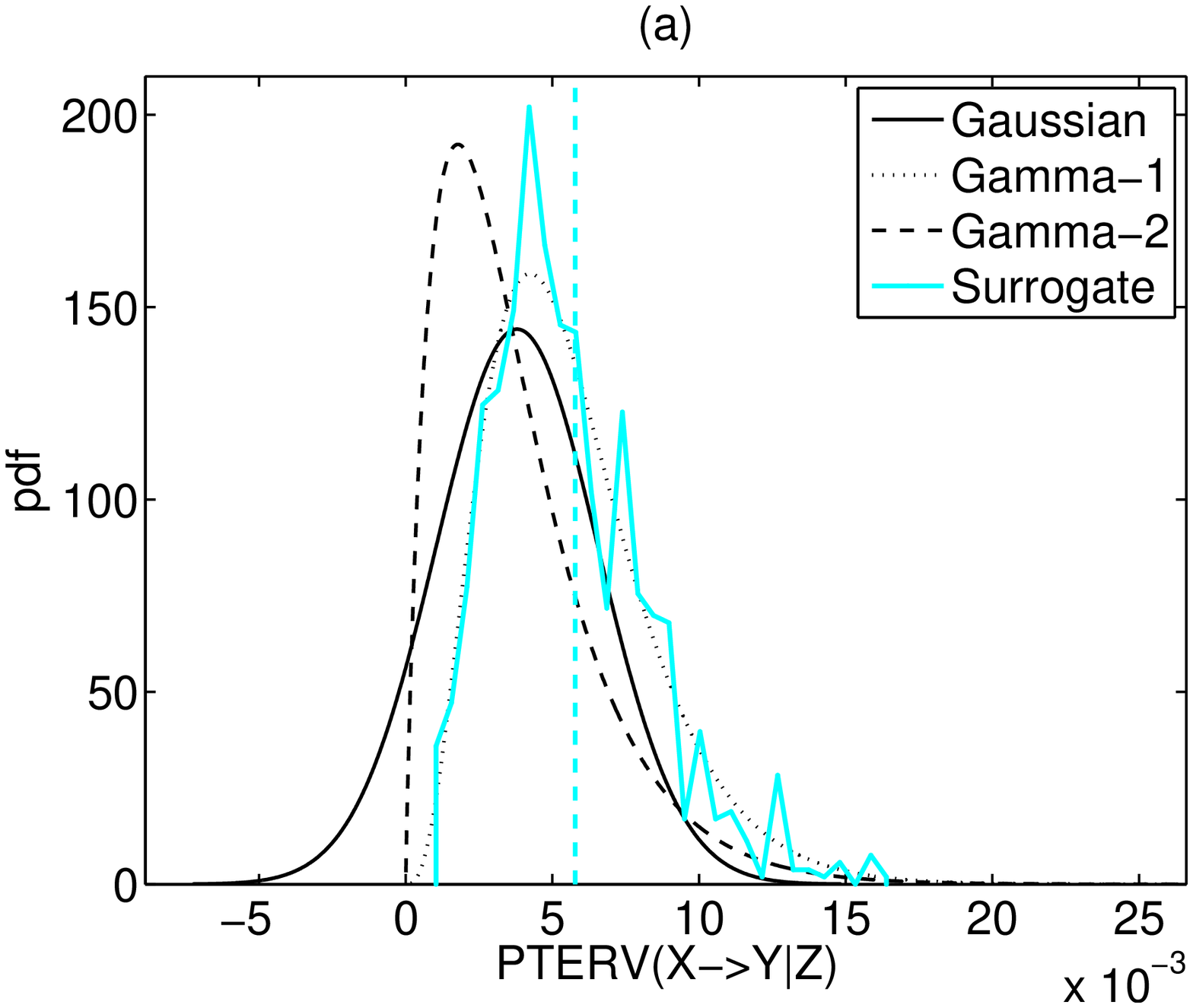}
\includegraphics[width=6cm]{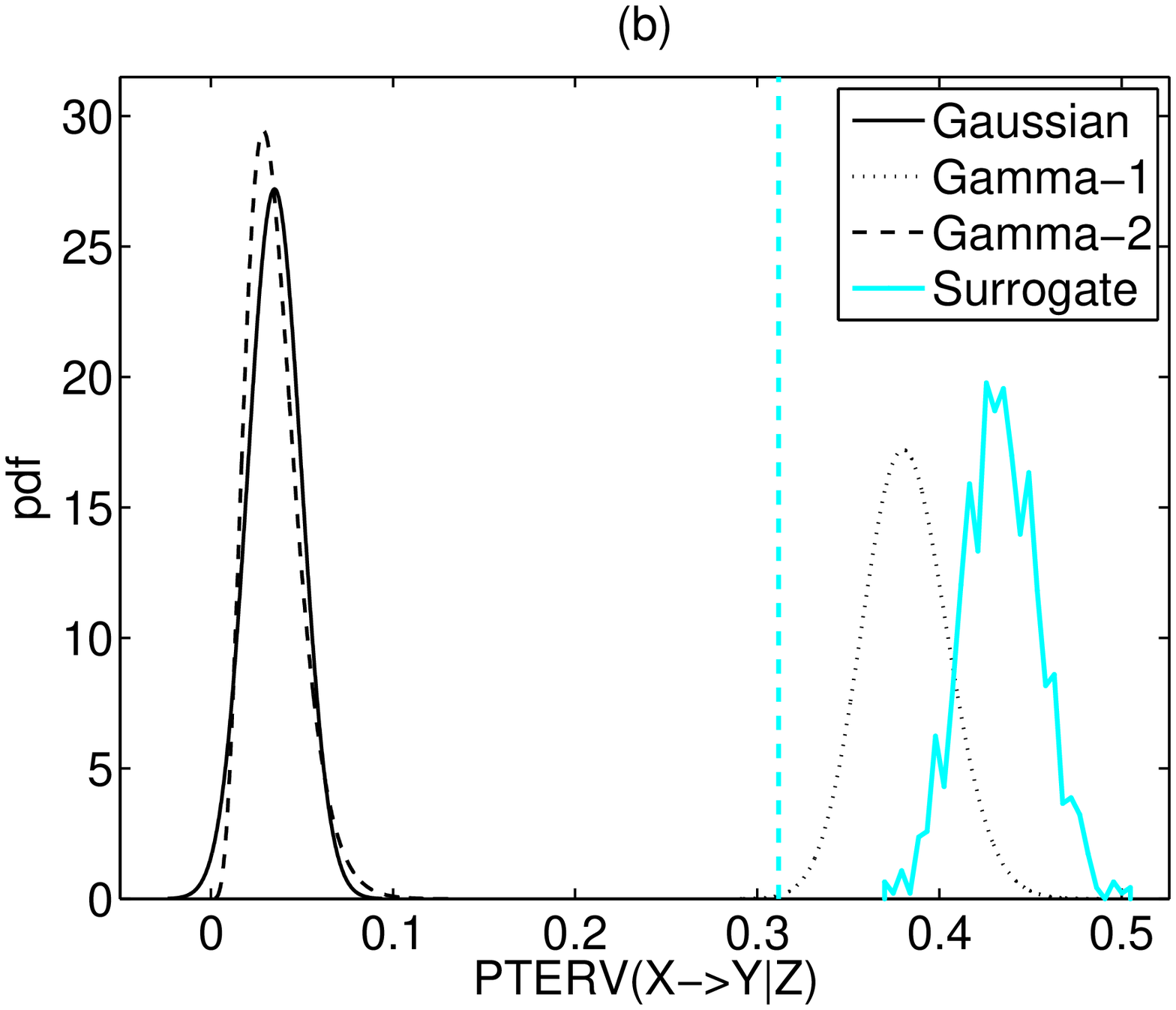}}
\hbox{\includegraphics[width=6cm]{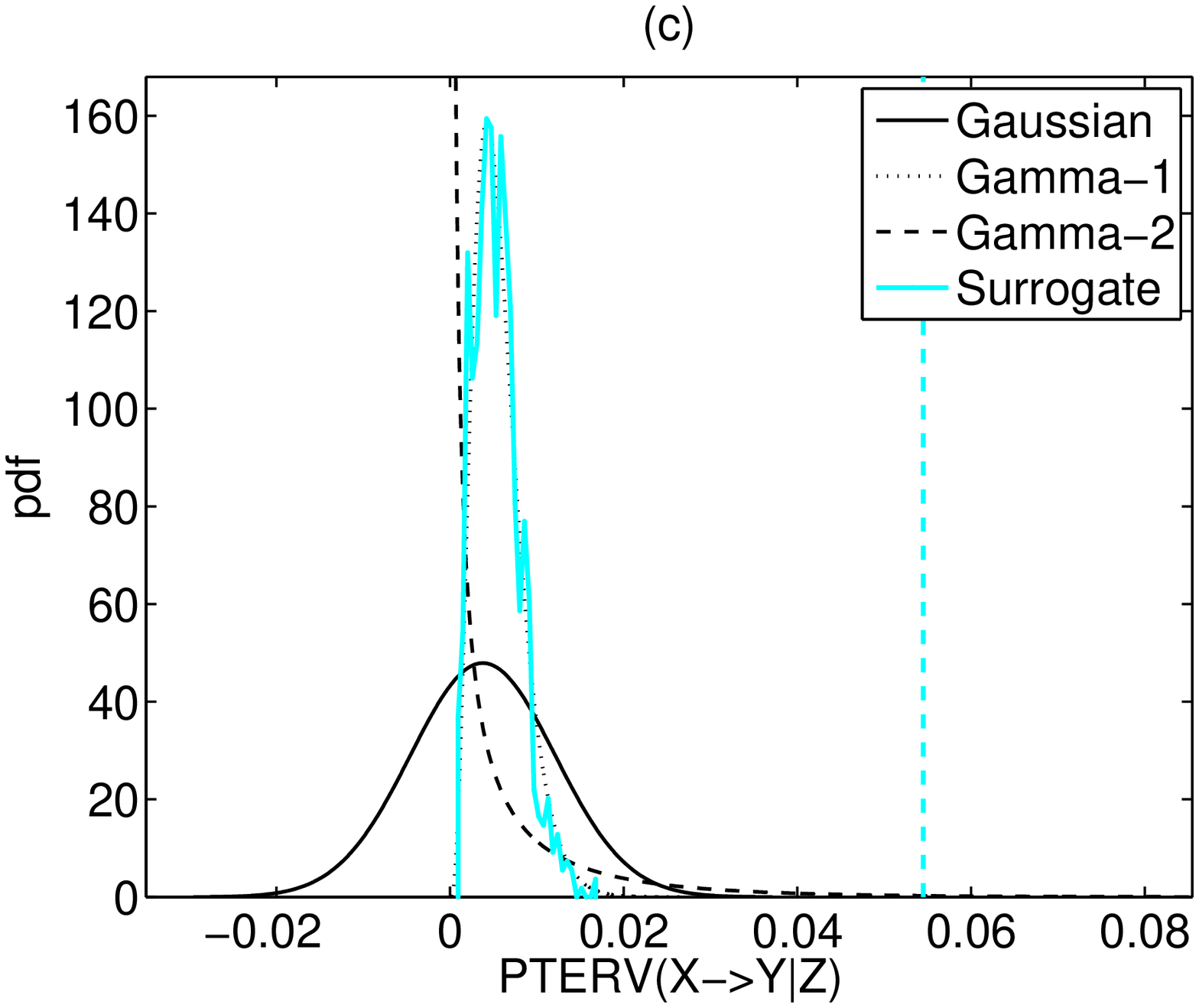}
\includegraphics[width=6cm]{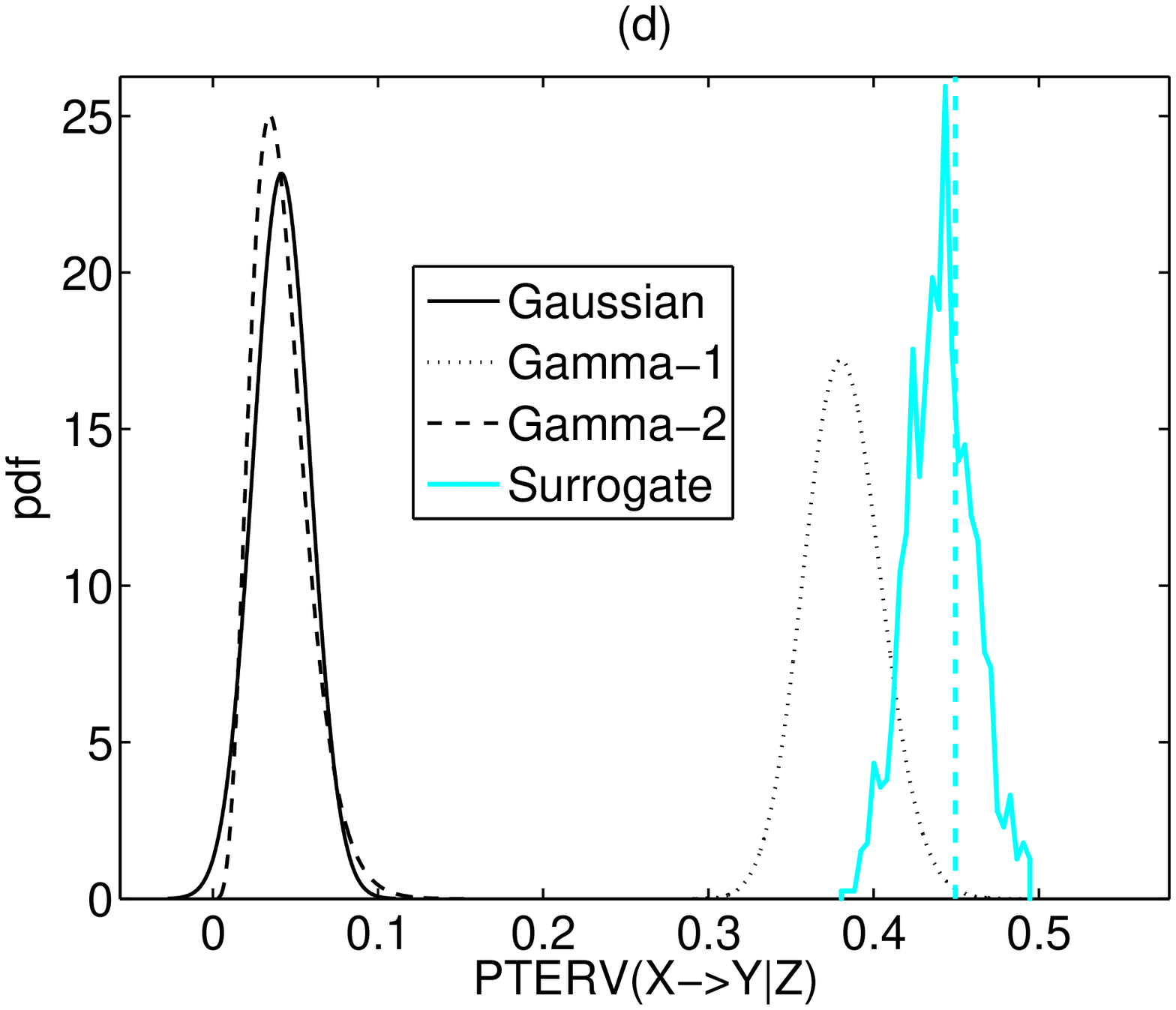}}
\hbox{\includegraphics[width=6cm]{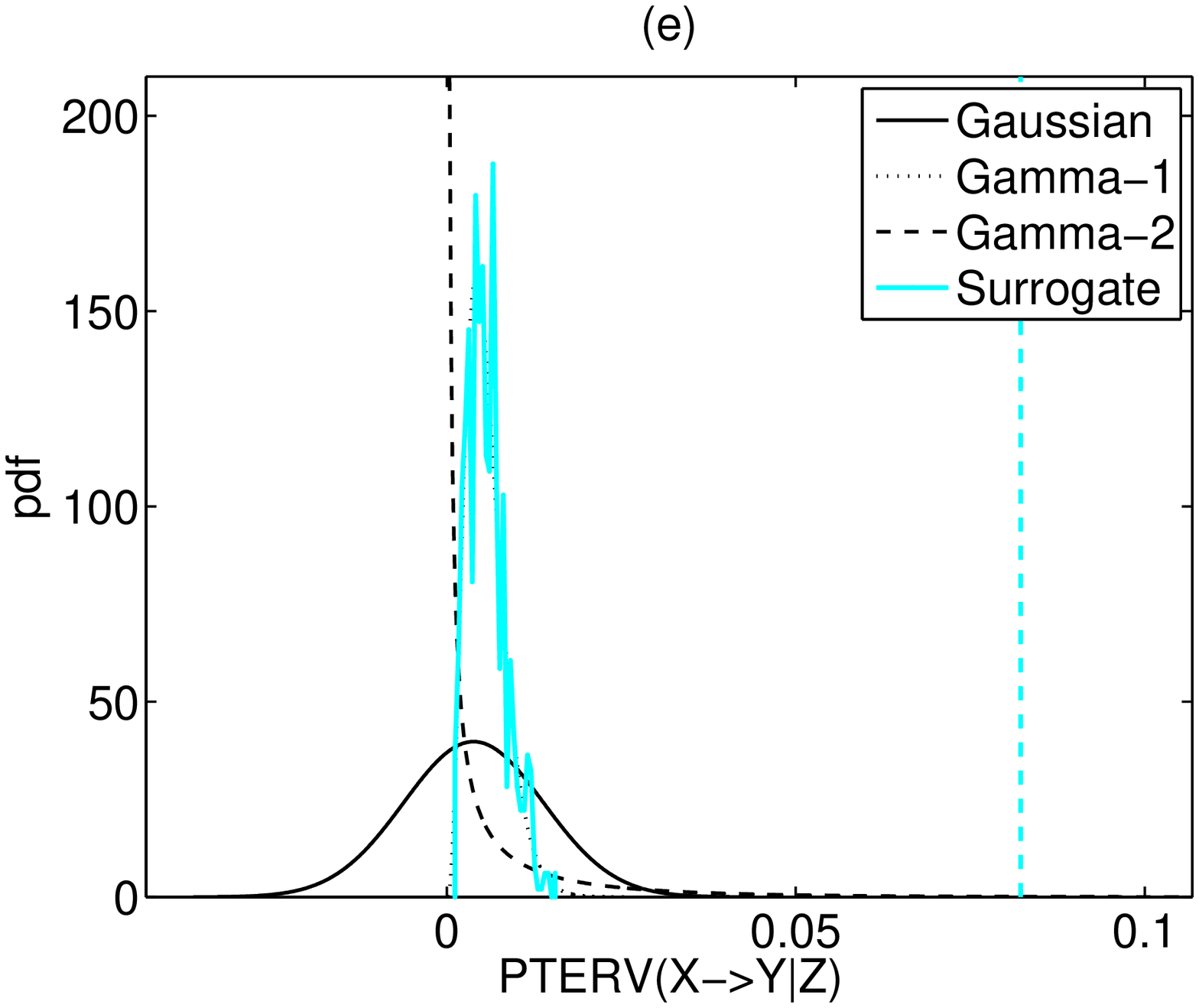}
\includegraphics[width=6cm]{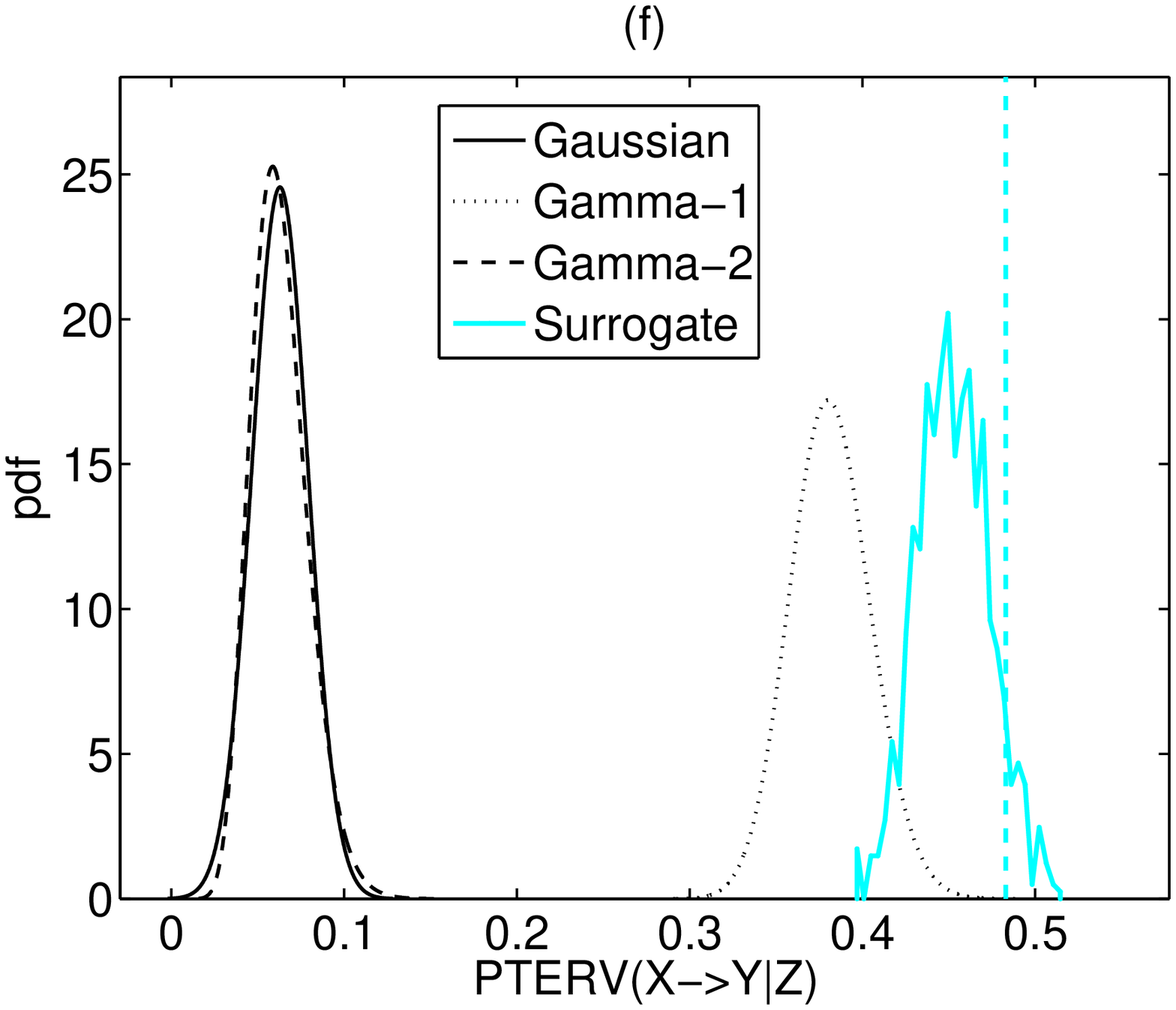}}
\caption{The three parametric approximations of the null
distribution of PTERV ($T=1$) and the distribution formed by 1000
time-shifted surrogates, as given in the legend, for one
realization of length $N=1024$ from the system in
eq.(\ref{eq:example1}) with $b=-1$, $d=0.8$. The other parameters
differ in the six panels as follows: (a) $a=2$, $c=0$, $m=2$, (b)
$a=2$, $c=0$, $m=3$, (c) $a=2$, $c=1$, $m=2$, (d) $a=2$, $c=1$,
$m=3$, (e) $a=0$, $c=1$, $m=2$, (f) $a=0$, $c=1$, $m=3$. The observed value of PTERV is shown by a vertical line.}
\label{fig:parsurdistrPTERV}       
\end{figure}

The randomization test is compared with the parametric tests on
the basis of the example in eq.(\ref{eq:example1}). For one
realization of the same system, the three parametric null
distributions and the null distribution formed by the PTERV
values from 1000 surrogates are shown in
Fig.~\ref{fig:parsurdistrPTERV}.
For $c=0$ the variables $X$ and $Y$ are conditionally independent
and H$_0$ holds. For this case and $m=2$, as shown in
Fig.~\ref{fig:parsurdistrPTERV}a, we obtain no rejection for any
of the four one-sided tests, where the test statistic is
$I_{\ssmall{obs}}(\hat{\mathbf{y}}^T_{t} ; \hat{\mathbf{x}}_t |
\hat{\mathbf{y}}_t,\hat{\mathbf{z}}_t)$ and the rejection regions
are formed from each approximate null distribution. In this case
the surrogate distribution is well matched by the Gamma-1
distribution. However, for $m=3$,
$I_{\ssmall{obs}}(\hat{\mathbf{y}}^T_{t} ; \hat{\mathbf{x}}_t |
\hat{\mathbf{y}}_t,\hat{\mathbf{z}}_t)$ lies well on the right of
the Gaussian and Gamma-2 distributions giving false rejection, and
on the left of the Gamma-1 and surrogate distributions giving
correctly no rejection (see Fig.~\ref{fig:parsurdistrPTERV}b).
The failure of the surrogate distribution to contain
$I_{\ssmall{obs}}(\hat{\mathbf{y}}^T_{t} ; \hat{\mathbf{x}}_t |
\hat{\mathbf{y}}_t,\hat{\mathbf{z}}_t)$ is attributed to
the inconsistency that the original $X$ variable depends on $Z$
but the surrogate $X$ does not.

When $X$ drives $Y$ ($c=1$), all four approaches give correctly
confident rejection of H$_0$ for $m=2$ (see
Fig.~\ref{fig:parsurdistrPTERV}c), but for $m=3$ the results are
the same as for $c=0$ (Fig.~\ref{fig:parsurdistrPTERV}d). Thus the
inconsistency mentioned above affects the power of the surrogate
data test, when there is dependence between the driving variable
and the conditioning variable. When we remove this dependence,
setting $a=0$, the surrogate data test performs well for both
$m=2$ and $m=3$ (see Fig.~\ref{fig:parsurdistrPTERV}e and f).
Indeed for $m=3$, the estimated probability of rejection of H$_0$
at the significance level $\alpha=0.05$ from 1000 realizations is
99.8 for the test with Gamma-1 approximation, 68.6 for the
randomization test and 1.0 for Gaussian and Gamma-2
approximations.

We argued in Sec.~\ref{subsec:PTERV} that PSTE is suboptimal for
its purpose and advocated for using PTERV instead. Having $\hat{\mathbf{y}}_{t+T}$ as
response variable  in PSTE (instead of
$\hat{\mathbf{y}}^T_{t}$ used in PTERV) introduces stronger
dependence of the response variable on the conditioning variable
because $\hat{\mathbf{y}}_{t+T}$ and $\hat{\mathbf{y}}_{t}$ are
formed from sample vectors that may have common components. In
turn this produces smaller values of CMI and thus it is harder to
identify significant causal effects. In particular, the power of
the randomization test with PSTE (quantified by the relative frequency of rejections of the null hypothesis of no coupling when there are true direct causal connections) is smaller even when there is no
dependence between the driving variable $X$ and the conditioning
variable $Z$. For example, for the last setting of the system in eq.(\ref{eq:example1}) ($a=0$,
$b=1$, $c=1$, $m=3$), using PSTE as test statistic there are no rejections for the randomization test
and the test with Gamma-1 approximation because PSTE lies always on
the left tail of both null distributions.

\section{Simulations and Results}
\label{sec:Simulations}

We assess the ability of PTERV to identify the correct direct causal effects in multivariate time series and compare it to PSTE and PTE (using the $k$-nearest neighbor estimate and $k=5$). For PTERV and PSTE we apply the parametric tests and the randomization test, and for PTE only the randomization test. To account for multiple testing, the correction of false discovery rate (FDR) is employed \cite{Benjamini95}. 
According to FDR, we order the $p$-values of the $K(K-1)$ tests, $p_1 \leq p_2 \leq ... \leq p_{K(K-1)}$, and if $p_k$ is the largest $p$-value for which $p_k \leq \alpha k /(K(K-1))$ holds, then all H$_0$ regarding $p_1,\ldots,p_k$ are rejected at the significance level $\alpha$. Thus the larger the $K$, the smaller the $p$-value has to be to give rejection. The implication of this for the randomization test is that the number $M$ of surrogates has to be large because the smallest $p$-value that can be obtained from rank ordering is roughly $1/(M+1)$.\footnote{Using the correction for the empirical cumulative function in \cite{Yu01}, we compute the $p$-value for the one-sided test as $1 - (r^0-0.326)/(M+1+0.348)$, where $r^0$ is the rank of the original measure value in the ordered list of $M+1$ values.} In the simulations we use $K$ up to 5 and setting $M=100$ we did not encounter any insufficiency of the randomization test due to the requirements of the FDR correction. 

\subsection{Stationary time series}
\label{subsec:stationary}

We first consider the stationary multivariate time series from the system of $K$ coupled H\'{e}non maps, defined as 
\begin{eqnarray*}
x_{i,t} & = & 1.4-x_{i,t-1}^2+0.3x_{i,t-2}, \quad   i=1,K  \\ 
x_{i,t} & = & 1.4-0.5C(x_{i-1,t-1}+x_{i+1,t-1})+(1-C)x_{i,t-1}^2+0.3x_{i,t-2}, \quad  i=2,\ldots,K-1
\end{eqnarray*}
where the parameter $C$ determines the strength of coupling. The results for weak coupling ($C=0.2$) and for $K=3$ and $K=5$ are shown in Table~\ref{tab:HenonK}.
\begin{table}
\begin{tabular}{lrrrrrrr} 
\hline\noalign{\smallskip}
& \multicolumn{3}{c}{PTERV} & \multicolumn{3}{c}{PSTE} & \multicolumn{1}{c}{PTE} \\ 
$K=3$ & Gaussian & Gamma-1 & Surrogate & Gaussian & Gamma-1 & Surrogate & Surrogate \\ 
\noalign{\smallskip}\hline\noalign{\smallskip}
$X_1 \rightarrow X_2$ & 94 & 81 & 94 & 32 & 6 & 7 & 100 \\ 
$X_2 \rightarrow X_1$ & 1 & 0 & 3 & 1 & 0 & 0 & 2 \\ 
$X_1 \rightarrow X_3$ & 0 & 0 & 2 & 0 & 0 & 1 & 6 \\ 
$X_3 \rightarrow X_1$ & 1 & 0 & 3 & 1 & 0 & 0 & 1 \\ 
$X_2 \rightarrow X_3$ & 0 & 0 & 2 & 0 & 0 & 1 & 5 \\ 
$X_3 \rightarrow X_2$ & 94 & 77 & 90 & 34 & 8 & 7 & 100 \\ 
\noalign{\smallskip}\hline\noalign{\smallskip} 
$K=5$ & & & & & & & \\ 
\noalign{\smallskip}\hline\noalign{\smallskip}
$X_1 \rightarrow X_2$ & 100 & 40 & 77 & 95 & 9 & 30 & 69 \\ 
$X_2 \rightarrow X_1$ & 13 & 0 & 2 & 13 & 0 & 1 & 3 \\ 
$X_1 \rightarrow X_3$ & 59 & 0 & 3 & 80 & 0 & 1 & 2 \\ 
$X_3 \rightarrow X_1$ & 14 & 0 & 4 & 8 & 0 & 2 & 4 \\ 
$X_1 \rightarrow X_4$ & 59 & 0 & 0 & 72 & 0 & 2 & 0 \\ 
$X_4 \rightarrow X_1$ & 14 & 0 & 0 & 9 & 0 & 0 & 1 \\ 
$X_1 \rightarrow X_5$ & 20 & 0 & 1 & 11 & 0 & 1 & 3 \\ 
$X_5 \rightarrow X_1$ & 10 & 0 & 2 & 9 & 0 & 2 & 0 \\ 
$X_2 \rightarrow X_3$ & 97 & 46 & 51 & 100 & 52 & 33 & 69 \\ 
$X_3 \rightarrow X_2$ & 98 & 32 & 39 & 98 & 32 & 23 & 67 \\ 
$X_2 \rightarrow X_4$ & 70 & 12 & 6 & 90 & 15 & 4 & 2 \\ 
$X_4 \rightarrow X_2$ & 66 & 10 & 3 & 85 & 13 & 3 & 2 \\ 
$X_2 \rightarrow X_5$ & 12 & 0 & 2 & 8 & 0 & 1 & 4 \\ 
$X_5 \rightarrow X_2$ & 54 & 0 & 0 & 67 & 1 & 0 & 3 \\ 
$X_3 \rightarrow X_4$ & 98 & 28 & 39 & 100 & 32 & 18 & 68 \\ 
$X_4 \rightarrow X_3$ & 96 & 48 & 52 & 99 & 46 & 35 & 67 \\ 
$X_3 \rightarrow X_5$ & 11 & 0 & 5 & 10 & 0 & 5 & 1 \\ 
$X_5 \rightarrow X_3$ & 68 & 0 & 3 & 79 & 0 & 1 & 5 \\ 
$X_4 \rightarrow X_5$ & 17 & 0 & 6 & 7 & 0 & 4 & 3 \\ 
$X_5 \rightarrow X_4$ & 100 & 50 & 69 & 98 & 13 & 21 & 69 \\ 
\noalign{\smallskip}\hline  
\end{tabular}
\caption{The number of rejections (using the FDR correction) of the parametric tests, denoted as Gaussian and Gamma-1, with PTERV and PSTE and the randomization tests, denoted as Surrogate, with PTERV, PSTE and PTE for 100 realizations of the system of $K=3$ and $K=5$ coupled H\'{e}non maps with $C=0.2$, $N=1024$. The other parameters are $m=2$, $\tau=1$ and $T=1$.}
\label{tab:HenonK}
\end{table}
The results for Gamma-2 test are very similar to those of Gaussian test and are not shown. For all non-existing or indirect couplings, all measures and with all tests give none or only few rejections using FDR at the level of significance $\alpha=0.05$ (5 out of 100 realizations), except the Gaussian approximation that gives always significant rejections for $K=5$. The randomization test with PTE detects best the true direct causal effects, followed closely by the randomization test with PTERV, while using PSTE the power of the randomization test is decreased for $K=5$ and even nullified for $K=3$. The parametric test with Gamma-1 approximation has smaller power than the randomization test using PTERV and the opposite is observed using PSTE (the latter only for $K=5$). 

\begin{figure}
\centering
\hbox{\includegraphics[width=6cm]{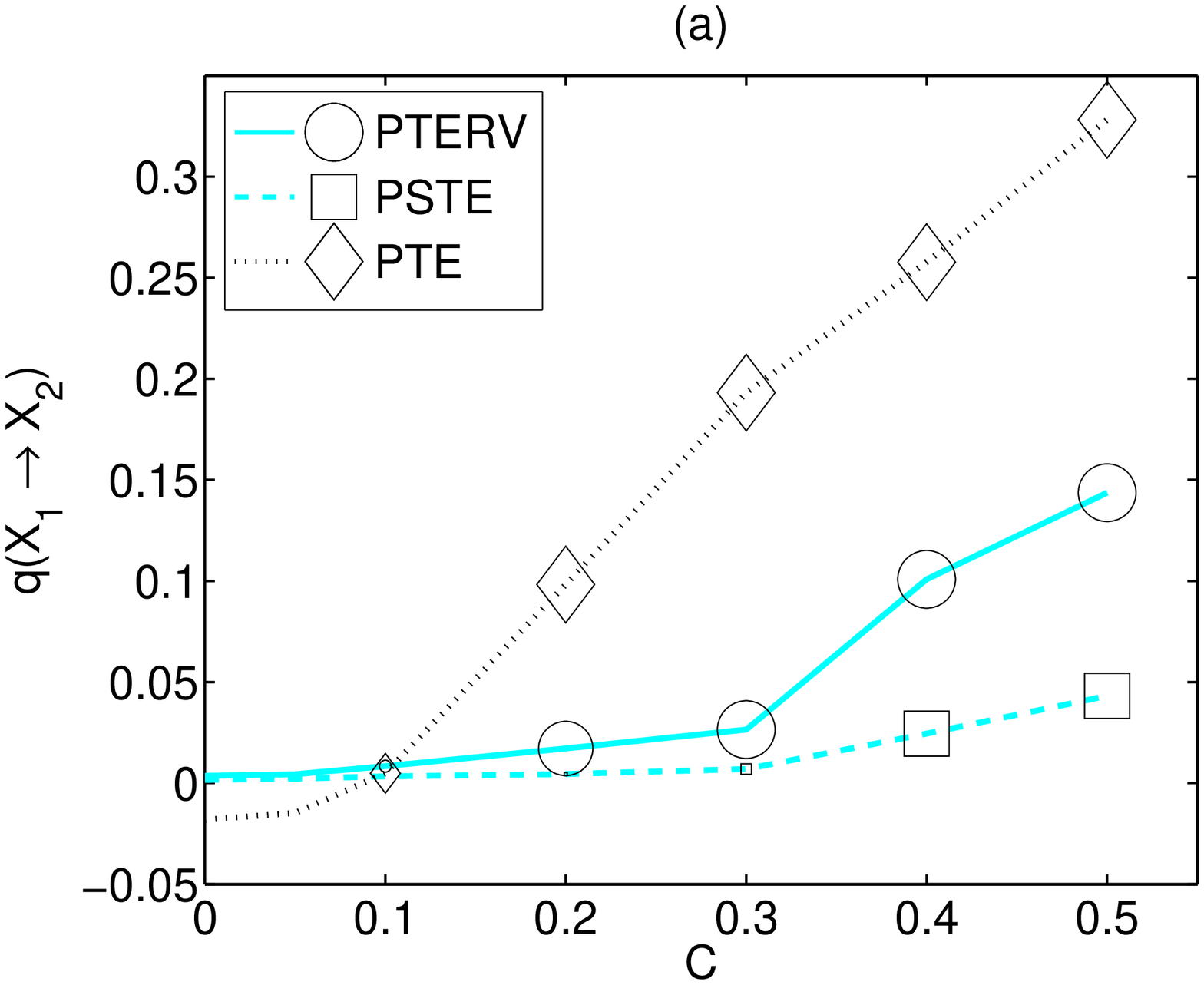}
\includegraphics[width=6cm]{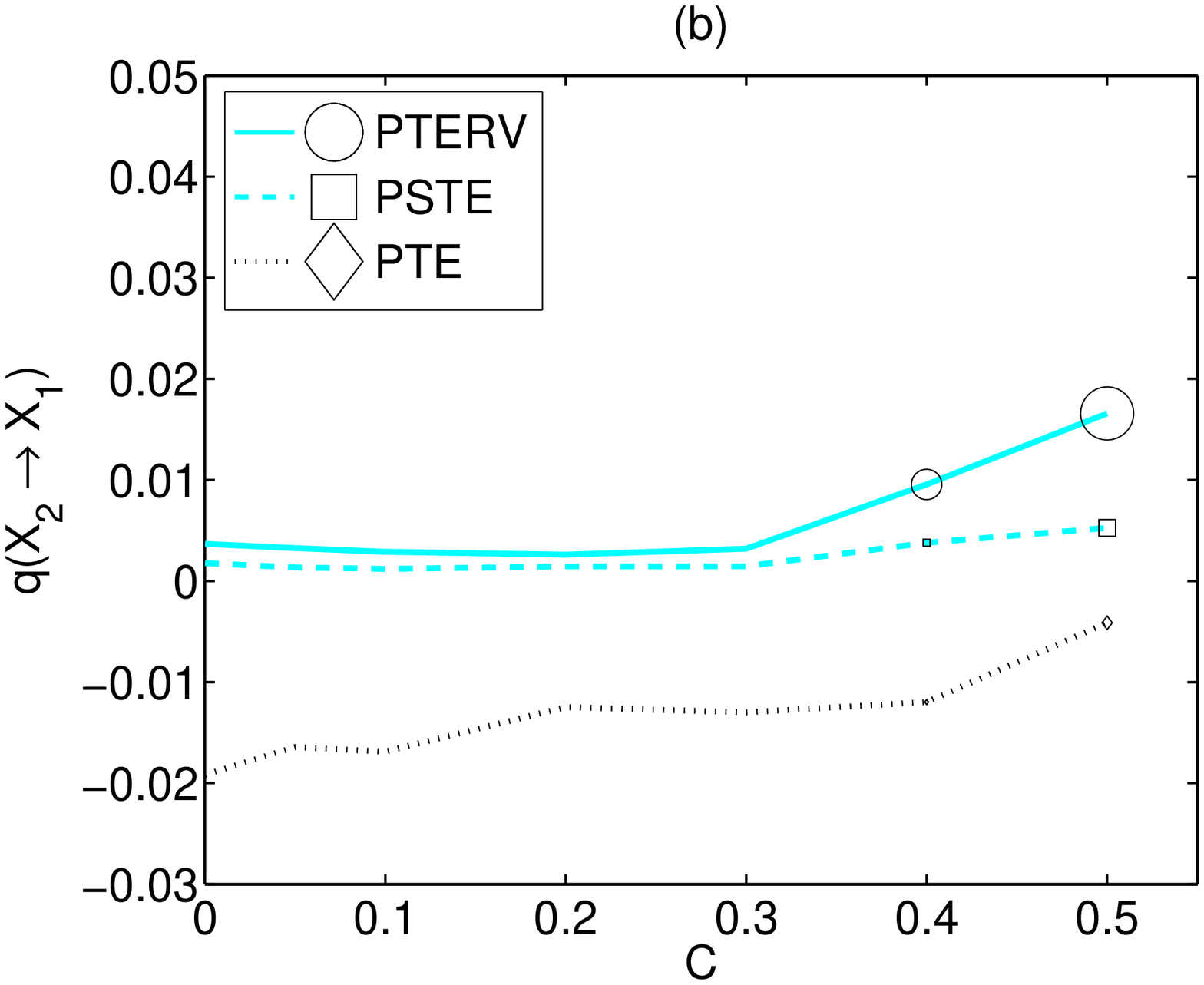}}
\caption{PTERV, PSTE and PTE measures (denoted collectively $q$) are given as functions of the coupling strength $C$ for the true direct causality $X_1 \rightarrow X_2$ in (a) and the non-existing coupling $X_2 \rightarrow X_1$ in (b) for the coupled H\'{e}non maps of 3 variables ($m=2$, $\tau=1$, $T=1$). The number of rejections in 100 realizations of the randomization test determines the size of a symbol displayed for each measure and $C$, where in the legend the size of the symbols regards 100 rejections.}
\label{fig:HenonKdiffC}       
\end{figure}

The extension of the simulations to different coupling strengths $C$ reveals the superiority of PTE and insufficiency of PSTE in detecting direct causal effects when the coupling is weak. In Fig.~\ref{fig:HenonKdiffC}, PTERV, PSTE and PTE as well as the number of rejections in 100 realizations of the randomization tests are drawn for a range of $C$.
Though the three measures are drawn at the same scale, their range is not comparable because PTE is estimated by nearest neighbors on the continuous-valued time series and its bias has different range than the bias of PTERV and PSTE, and the bias of PTERV and PSTE also differ due to different number of states of $\hat{\mathbf{y}}_t^T$ and $\hat{\mathbf{y}}_{t+T}$, respectively. However, for the true coupling in Fig.~\ref{fig:HenonKdiffC}a, one can note that PTE tends to increase steeper than PTERV, which in turn increases steeper than PSTE. Accordingly, PTE and PTERV are likely to be found significant for as small $C$ as 0.1, while PSTE gets significant only for strong coupling ($C \geq 0.4$). The lower level of PTE is negative, so that zero PTE is actually found significant (see for $C=0.1$ in Fig.~\ref{fig:HenonKdiffC}a), which is counterintuitive. There are even some few rejections of H$_0$ of no coupling when PTE is negative, as shown for large $C$ for the erroneous causality $X_2 \rightarrow X_1$ in Fig.~\ref{fig:HenonKdiffC}b. For the latter setting, PTERV is found statistically significant showing the tendency to detect coupling in the wrong direction when the coupling strength is large.

For the simulation setting above, it was found that the Gaussian and Gamma-2 parametric tests have decreased power  compared to the Gamma-1 parametric test and the randomization test. Further simulations showed that this holds only for $m=2$ and for larger $m$ the bias is underestimated giving rejections for the Gaussian and Gamma-2 parametric tests, and on the other hand the Gamma-1 parametric test tends to be overly conservative. For example for $K=3$, $C=0.2$ and $m=3$, Gaussian and Gamma-2 parametric tests give always rejection and Gamma-1 parametric test gives always no rejection for any pair of variables. On the other hand, the randomization test gives optimal results for both PTERV and PSTE, i.e. 100 rejections when there is true direct causality and no rejections when there is not, and the same is obtained with PTE. We note also that PTERV and PSTE converge in their performance as $m$ gets larger because then the effect of defining the response rank vector differently weakens.  

We assess the measures on a system of three coupled Lorenz subsystems
\begin{equation}
\begin{array}{rcl}
\dot{x}_{1} & = & -10x_{1} + 10y_{1} \\
\dot{y}_{1} & = & -x_{1}z_{1} + 28x_{1} - y_{1} \\
\dot{z}_{1} & = & x_{1}y_{1} -8/3z_{1} 
\end{array}
\quad\quad
\begin{array}{rcl}
\dot{x}_{i} & = & -10x_{i} + 10y_{i} + C(x_{i-1}-x_{i}) \\
\dot{y}_{i} & = & -x_{i}z_{i} + 28x_{i} - y_{i} \\
\dot{z}_{i} & = & x_{i}y_{i} -8/3z_{i} 
\end{array}
\quad i=2,3
\label{eq:LorenzK}
\end{equation}

The system is solved using the explicit Runge-Kutta (4,5) method implemented in the solver {\tt ode45} in Matlab and the time series are generated at a sampling time of 0.01 time units. We observe the first variable of each of the three systems, i.e. the tuple of observed variables $(X,Y,Z)$ is assigned to $(x_{1},x_{2},x_{3})$, so the direct couplings are $X \rightarrow Y$ and $Y \rightarrow Z$. We use the same coupling strength $C$ for both couplings. For the flow system, longer time series are required to detect the information flow and in Table~\ref{tab:LorenzK} we show the results for $N=4096$ and $N=16384$ for weak coupling ($C=2$).
\begin{table}
\begin{tabular}{lrrrrr} 
\hline\noalign{\smallskip}
& \multicolumn{2}{c}{PTERV} & \multicolumn{2}{c}{PSTE} & \multicolumn{1}{c}{PTE} \\ 
$N=4096$ & Gamma-1 & Surrogate & Gamma-1 & Surrogate & Surrogate \\
\noalign{\smallskip}\hline\noalign{\smallskip}
$X \rightarrow Y$ & 0 & 11 & 0 & 9 & 100 \\ 
$Y \rightarrow X$ & 0 & 0 & 0 & 0 & 3 \\ 
$X \rightarrow Z$ & 0 & 1 & 0 & 1 & 6 \\ 
$Z \rightarrow X$ & 0 & 0 & 0 & 0 & 2 \\ 
$Y \rightarrow Z$ & 0 & 3 & 0 & 2 & 100 \\ 
$Z \rightarrow Y$ & 0 & 1 & 0 & 1 & 2 \\ 
\noalign{\smallskip}\hline\noalign{\smallskip} 
$N=16384$ & & & & & \\ 
\noalign{\smallskip}\hline\noalign{\smallskip}
$X \rightarrow Y$ & 0 & 99 & 0 & 92 & 100 \\ 
$Y \rightarrow X$ & 0 & 0 & 0 & 0 & 9 \\ 
$X \rightarrow Z$ & 0 & 2 & 0 & 3 & 43 \\ 
$Z \rightarrow X$ & 0 & 1 & 0 & 1 & 8 \\ 
$Y \rightarrow Z$ & 0 & 90 & 0 & 77 & 100 \\ 
$Z \rightarrow Y$ & 0 & 30 & 0 & 29 & 7 \\ 
\noalign{\smallskip}\hline  
\end{tabular}
\caption{The number of rejections (using the FDR correction) of the Gamma-1 parametric test with PTERV and PSTE and the randomization (surrogate) test with PTERV, PSTE and PTE for 100 realizations of the $K=3$ coupled Lorenz systems with $C=2$, for $N=4096$ and $N=16384$. The other parameters are $m=4$, $\tau=1$ and $T=1$.}
\label{tab:LorenzK}
\end{table}
Again the parametric tests fail for both PTERV and PSTE. For $N=4096$, the randomization tests with PTERV and PSTE have good significance (we use again $\alpha=0.05$ and FDR correction) but almost no power, while the randomization test with PTE has optimal power and significance. It seems that the rank measures are more data demanding and as we increased $N$ we could improve the power of their randomization test, as shown in Table~\ref{tab:LorenzK} for $N=16384$. However, both PTERV and PSTE tend to falsely detect the coupling $Z \rightarrow X$, while PTE detects also falsely and with higher estimated probability the indirect coupling $X \rightarrow Z$. 
Note that for this setting having a larger embedding dimension ($m=4$) PSTE performs as well as PTERV. 

Further simulations for different coupling strengths reveal shortcomings of all measures. 
\begin{figure}
\centering
\hbox{\includegraphics[width=6cm]{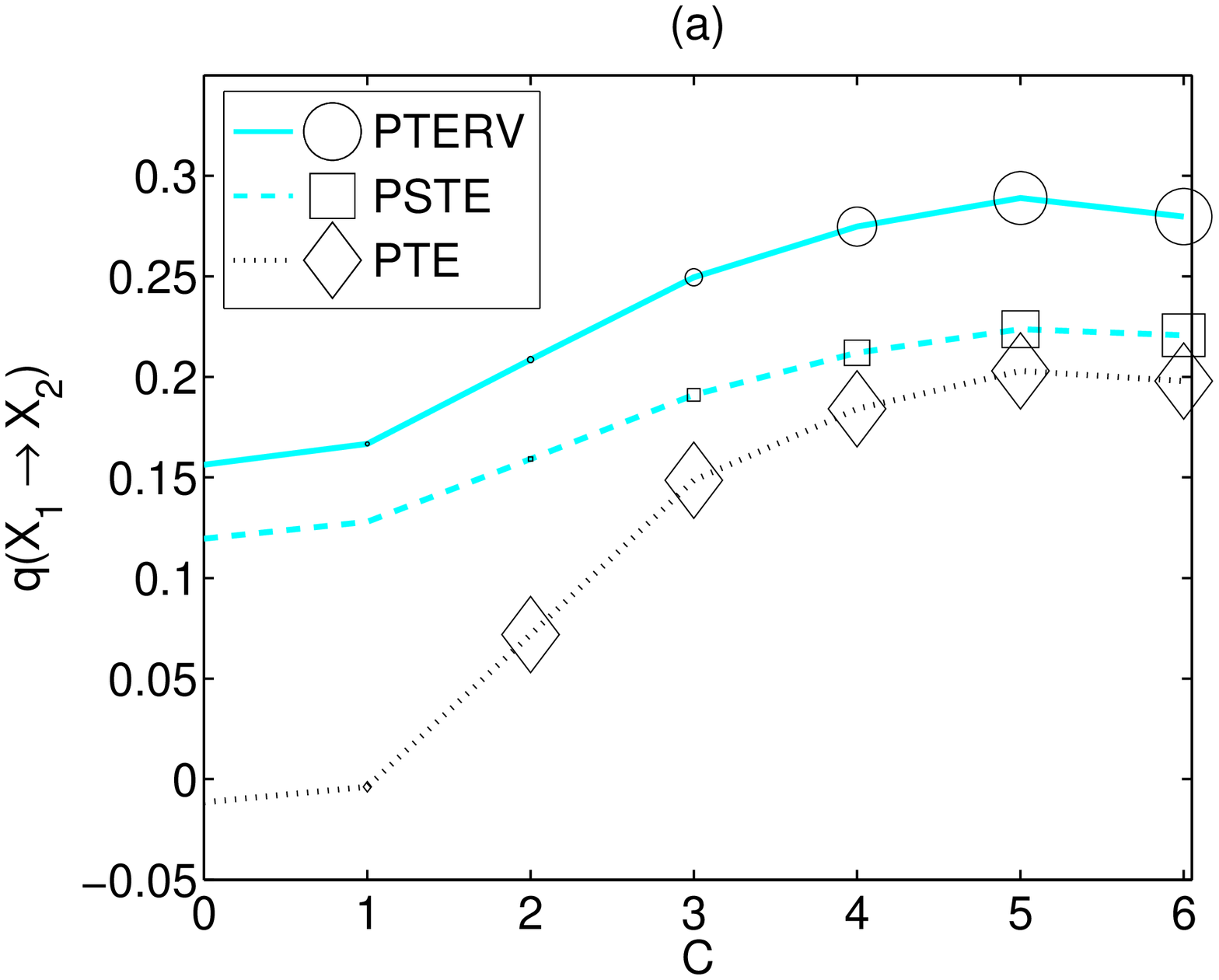}
\includegraphics[width=6cm]{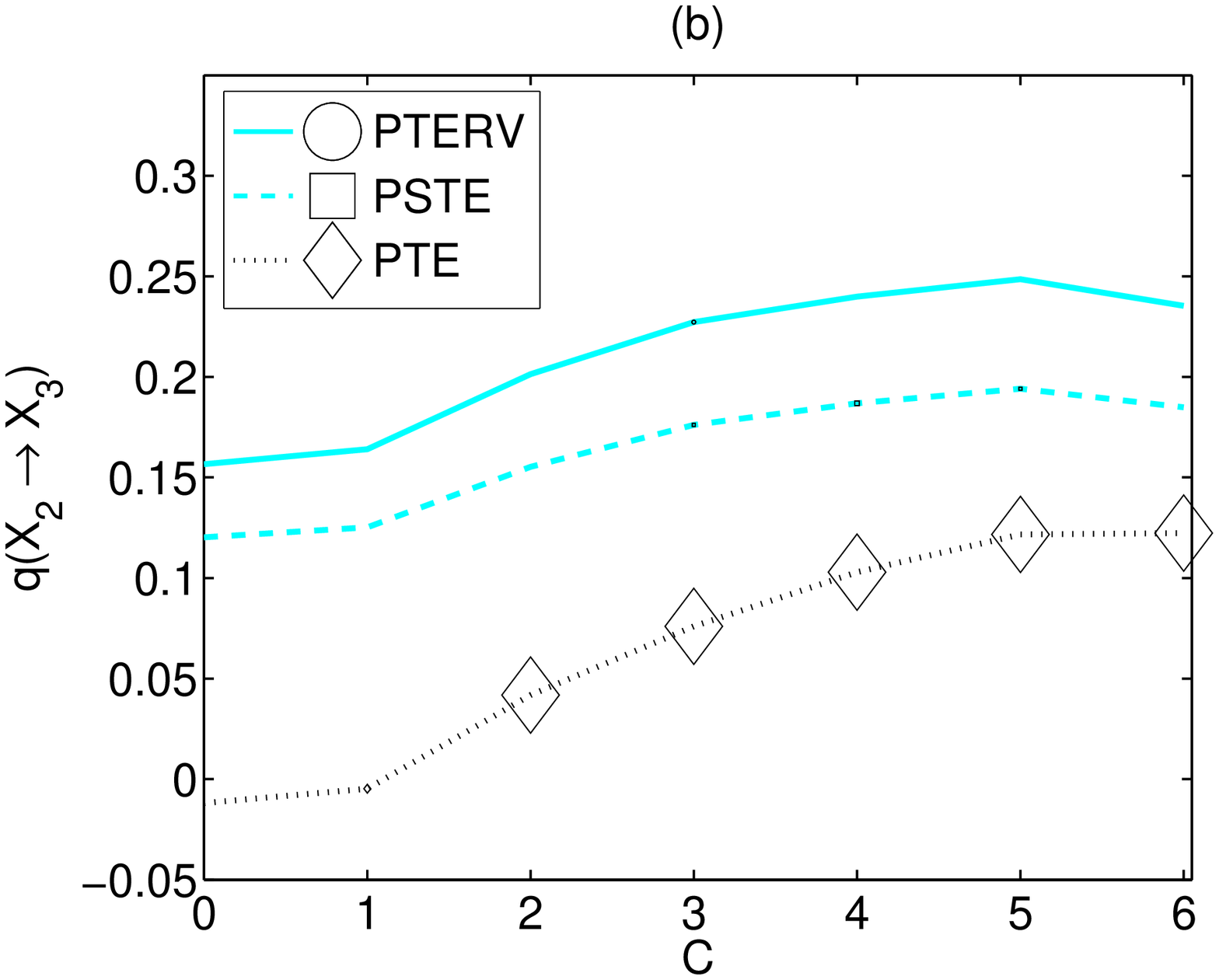}}
\hbox{\includegraphics[width=6cm]{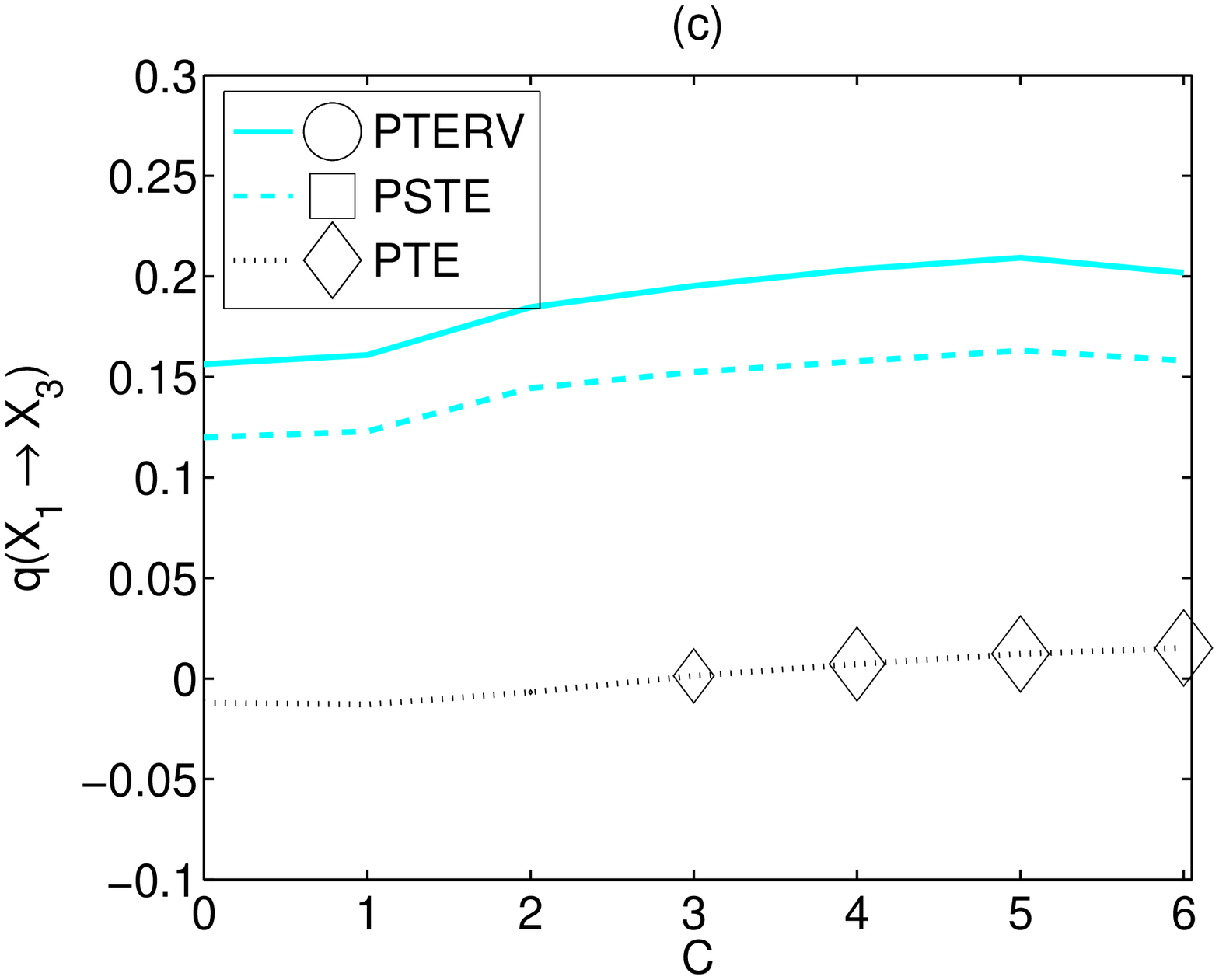}
\includegraphics[width=6cm]{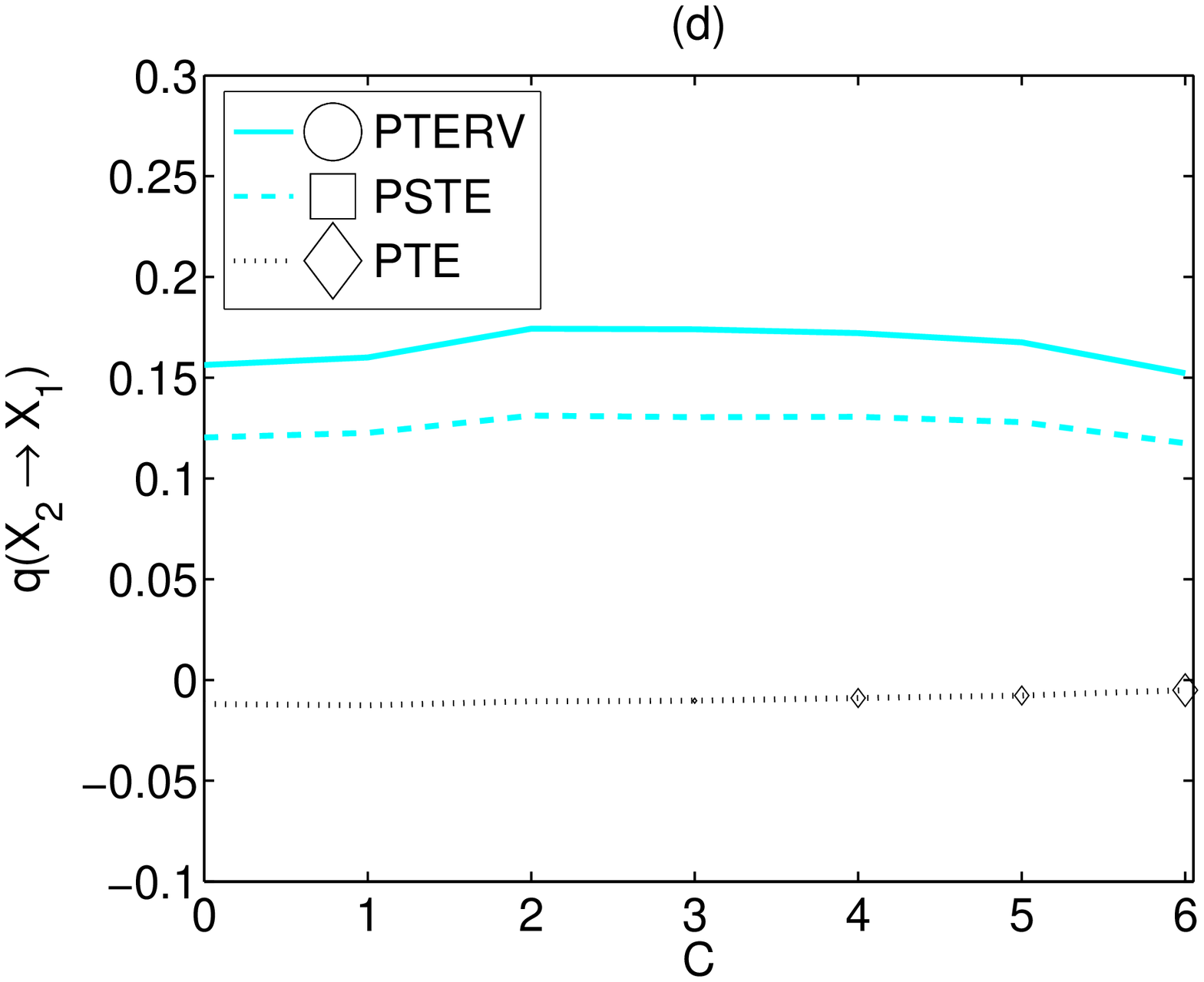}}
\caption{As Fig.~\ref{fig:HenonKdiffC} but for the three coupled Lorenz systems, $m=4$, $\tau=1$, $T=1$, $4096$, and the couplings $X \rightarrow Y$ in (a), $Y \rightarrow Z$ in (b), $X \rightarrow Z$ in (c), and $Y \rightarrow X$ in (d).}
\label{fig:LorenzKdiffC}       
\end{figure}
As shown in Fig.~\ref{fig:LorenzKdiffC} for $N=4096$, PTERV and PSTE increase with $C$ in the same way as PTE for the two true direct couplings. However, the power of the randomization tests for the two rank measures is much smaller than for PTE for the first true direct coupling ($X \rightarrow Y$ in Fig.~\ref{fig:LorenzKdiffC}a) and at the zero level for the second true direct coupling ($Y \rightarrow Z$ in Fig.~\ref{fig:LorenzKdiffC}b). On the other hand, both PTERV and PSTE are at the zero level when there is indirect or no coupling and the randomization test has the correct significance, unlike the randomization test with PTE that tends to reject with larger probability than the significance level, which increases with $C$. The latter is more pronounced for the indirect coupling ($X \rightarrow Z$ in Fig.~\ref{fig:LorenzKdiffC}c), while for no coupling, significant rejections are observed for large $C$ (as for $Y \rightarrow X$ shown in Fig.~\ref{fig:LorenzKdiffC}d).

Other simulations not shown here suggest that it is difficult to estimate the correct causality structure in this system and the results are sensitive to the choice of the parameters $m$, $\tau$ and $T$. We note that when we added observational noise to the time series of the H\'{e}non and Lorenz coupled system we observed that the power of the randomization tests were reduced accordingly and in the same way for all causality measures.

\subsection{Non-stationary time series}
\label{subsec:nonstationary}

The simulation results showed that PTERV cannot reach the performance of PTE in detecting correctly the direct causal effects. However, there is a practical situation that PTERV can indeed be more useful than PTE, namely in the presence of drifts in the time series, a common data condition in many applications. PTE along with many other measures of Granger causality assume stationarity of the time series. On the other hand, the measures applied to the ranks rather than the samples are little affected by non-stationarity in the mean. This is so because in the estimation of the effect of the driving variable on the response variable the level of their magnitudes is disregarded and only the relative magnitude ordering counts. 

\begin{figure}
\centering
\hbox{\includegraphics[width=6cm]{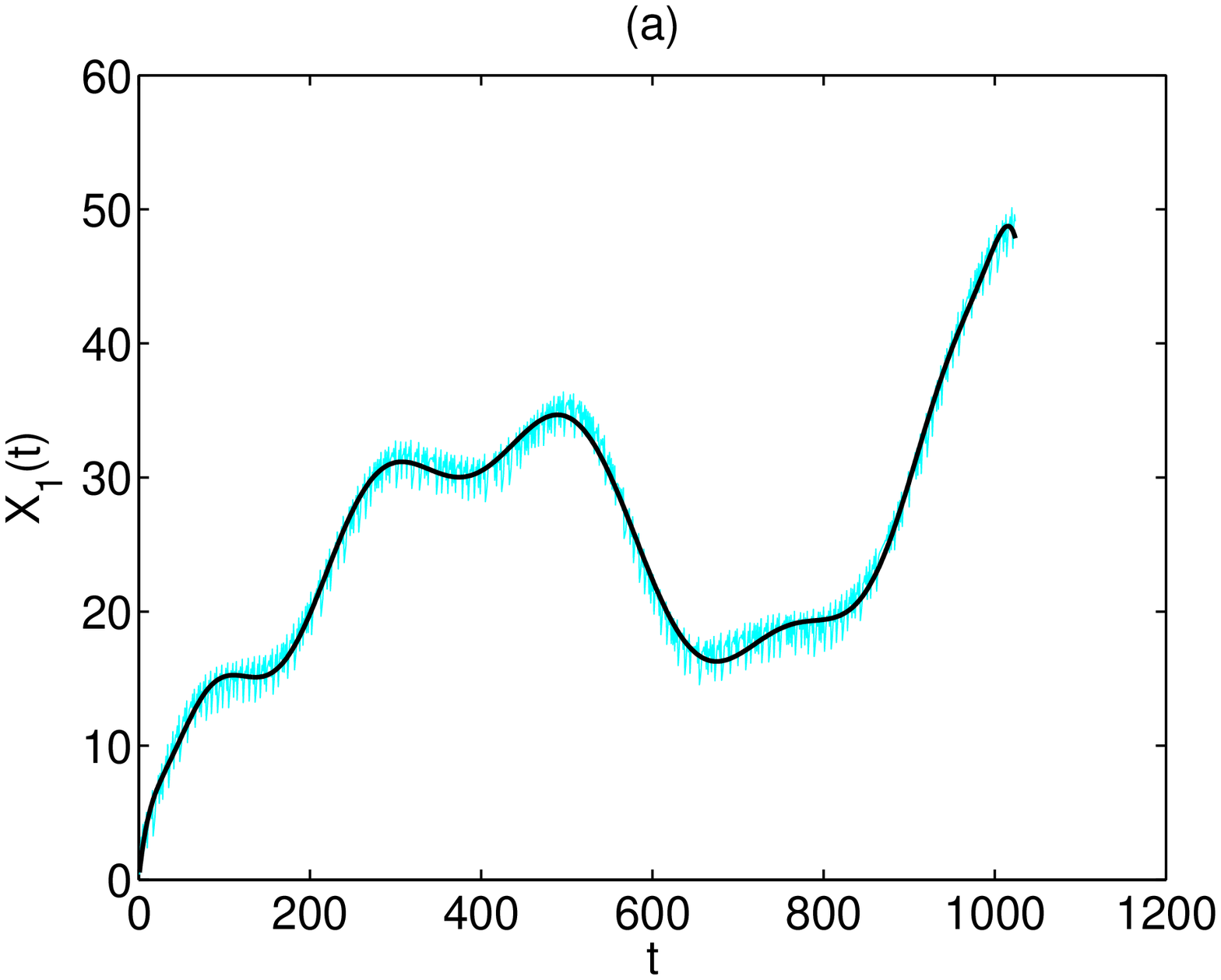}
\includegraphics[width=6cm]{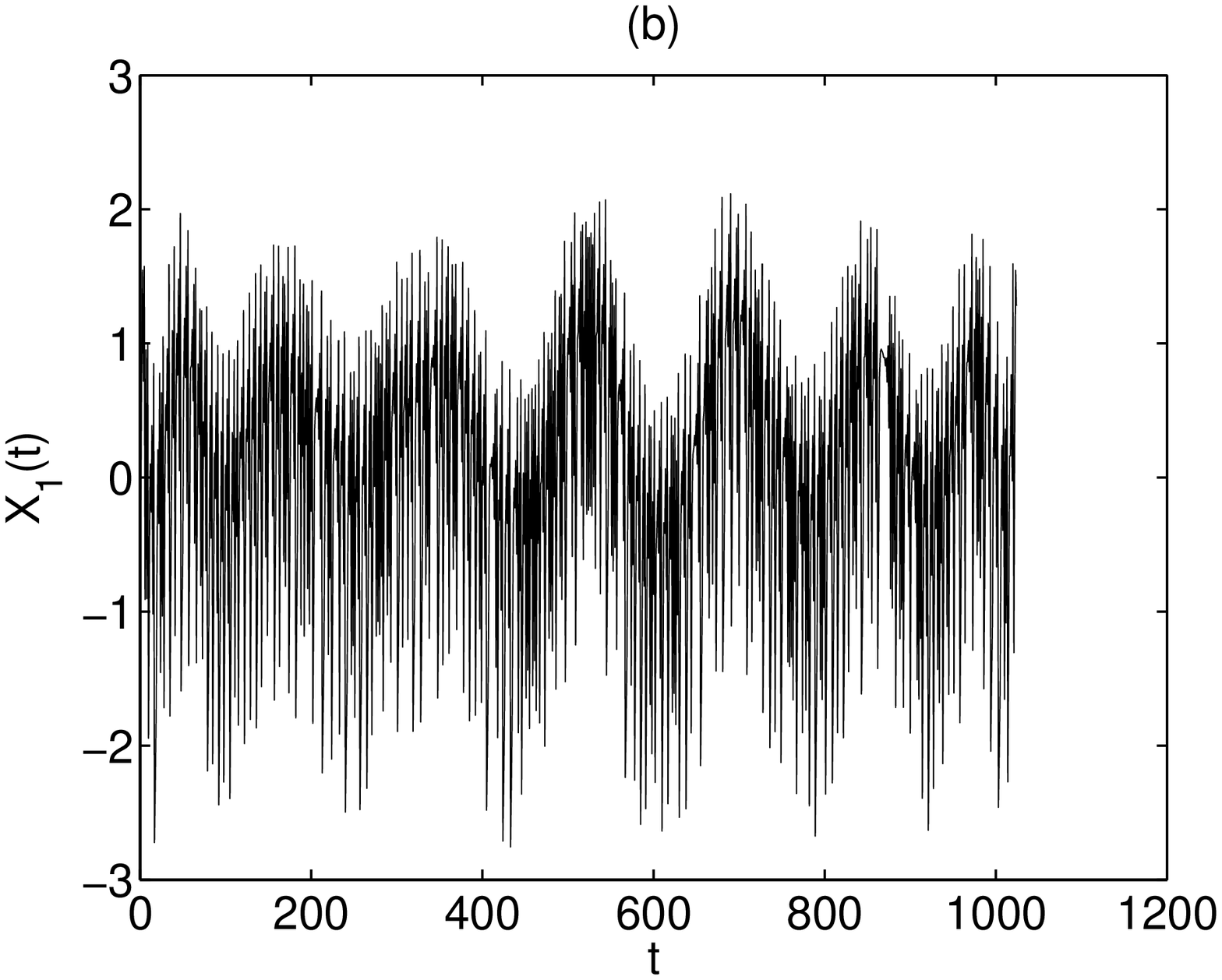}}
\caption{(a) A time series of the variable $X_1$ of the coupled H\'{e}non maps ($K=3$) after adding to it a stochastic trend. Superimposed is the fit of a polynomial of degree 15. (b) The time series in (a) detrended by subtracting the polynomial fit.}
\label{fig:detrending}       
\end{figure}

We demonstrate the appropriateness of PTERV as compared to PTE on synthetically constructed non-stationary time series. We assume the coupled H\'{e}non maps in three variables and add to each time series a stochastic trend that is computed as follows. First, a Gaussian random walk time series of the same length as the original time series is generated, where the SD of the random steps is the same as the SD of the H\'{e}non time series. Then a moving average smoothing of order 100 is applied to the random walk time series. This time series of smoothed stochastic trend is added to the time series of the first variable, $X_1$, of the coupled H\'{e}non maps, and the same process is repeated for $X_2$ and $X_3$. An example is shown in Fig.~\ref{fig:detrending}.
There are no standard methods to remove stochastic trends and depending on the employed approach different time series may be derived and the original dynamics may be altered \cite{Theiler93b}. For example, the fit of a polynomial of order 15 seems to match well the slow drift (Fig.~\ref{fig:detrending}a), but the derived time series after detrending, seems to have variations not expected in the original time series generated by the coupled H\'{e}non map (Fig.~\ref{fig:detrending}b).  

We compute the measures PTERV, PSTE and PTE and conduct the Gamma-1 parametric test and randomization test on 100 non-stationary multivariate time series from the coupled H\'{e}non maps with $C=0.2$, and also on the detrended time series subtracting the polynomial fit (see Table~\ref{tab:HenonKnonstationary}).
\begin{table}
\begin{tabular}{lrrrrr} 
\hline\noalign{\smallskip}
& \multicolumn{2}{c}{PTERV} & \multicolumn{2}{c}{PSTE} & \multicolumn{1}{c}{PTE} \\ 
& Gamma-1 & Surrogate & Gamma-1 & Surrogate & Surrogate \\ 
\noalign{\smallskip}\hline\noalign{\smallskip}
\multicolumn{6}{c}{time series with slow drifts} \\
\noalign{\smallskip}\hline\noalign{\smallskip}
$X_1 \rightarrow X_2$ & 56 & 67 & 8 & 13 & 1 \\ 
$X_2 \rightarrow X_1$ & 3 & 2 & 5 & 3 & 4 \\ 
$X_1 \rightarrow X_3$ & 2 & 3 & 3 & 4 & 0 \\ 
$X_3 \rightarrow X_1$ & 4 & 3 & 5 & 5 & 2 \\ 
$X_2 \rightarrow X_3$ & 1 & 3 & 0 & 3 & 2 \\ 
$X_3 \rightarrow X_2$ & 56 & 67 & 14 & 13 & 3 \\ 
\noalign{\smallskip}\hline\noalign{\smallskip} 
\multicolumn{6}{c}{time series after detrending} \\ 
\noalign{\smallskip}\hline\noalign{\smallskip} 
$X_1 \rightarrow X_2$ & 81 & 89 & 8 & 13 & 100 \\ 
$X_2 \rightarrow X_1$ & 0 & 1 & 0 & 0 & 16 \\ 
$X_1 \rightarrow X_3$ & 1 & 3 & 1 & 2 & 9 \\ 
$X_3 \rightarrow X_1$ & 0 & 2 & 1 & 2 & 6 \\ 
$X_2 \rightarrow X_3$ & 0 & 3 & 0 & 2 & 18 \\ 
$X_3 \rightarrow X_2$ & 76 & 89 & 8 & 14 & 100 \\ 
\noalign{\smallskip}\hline  
\end{tabular}
\caption{The number of rejections (using the FDR correction) of the Gamma-1 parametric test with PTERV and PSTE and the randomization (surrogate) tests with PTERV, PSTE and PTE for 100 realizations of $K=3$ coupled H\'{e}non maps with $C=0.2$, $N=1024$. The other parameters are $m=2$, $\tau=1$ and $T=1$. The upper part is for the non-stationary time series (after a smoothed slow drift is added) and the lower part for the time series obtained after detrending with a polynomial of order 15.}
\label{tab:HenonKnonstationary}
\end{table}
For the non-stationary time series, as expected, PTE fails to detect any coupling. On the other hand, PTERV and PSTE perform similarly to the case of no drifts with somehow smaller power of the randomization and Gamma-1 tests (compare Table~\ref{tab:HenonKnonstationary} to Table~\ref{tab:HenonK}). Again PSTE has no power to detect the true direct causal effects. When PTE is applied to the detrended time series, it regains detection of the true direct causal effects, but it falsely estimates also the opposite causal effects with an estimated probability close to 0.2. On the other hand, the parametric and randomization test with PTERV gains almost the same power as if there was no drift to be detrended, while the tests with PSTE show again no power. 

\begin{figure}
\centering
\hbox{\includegraphics[width=6cm]{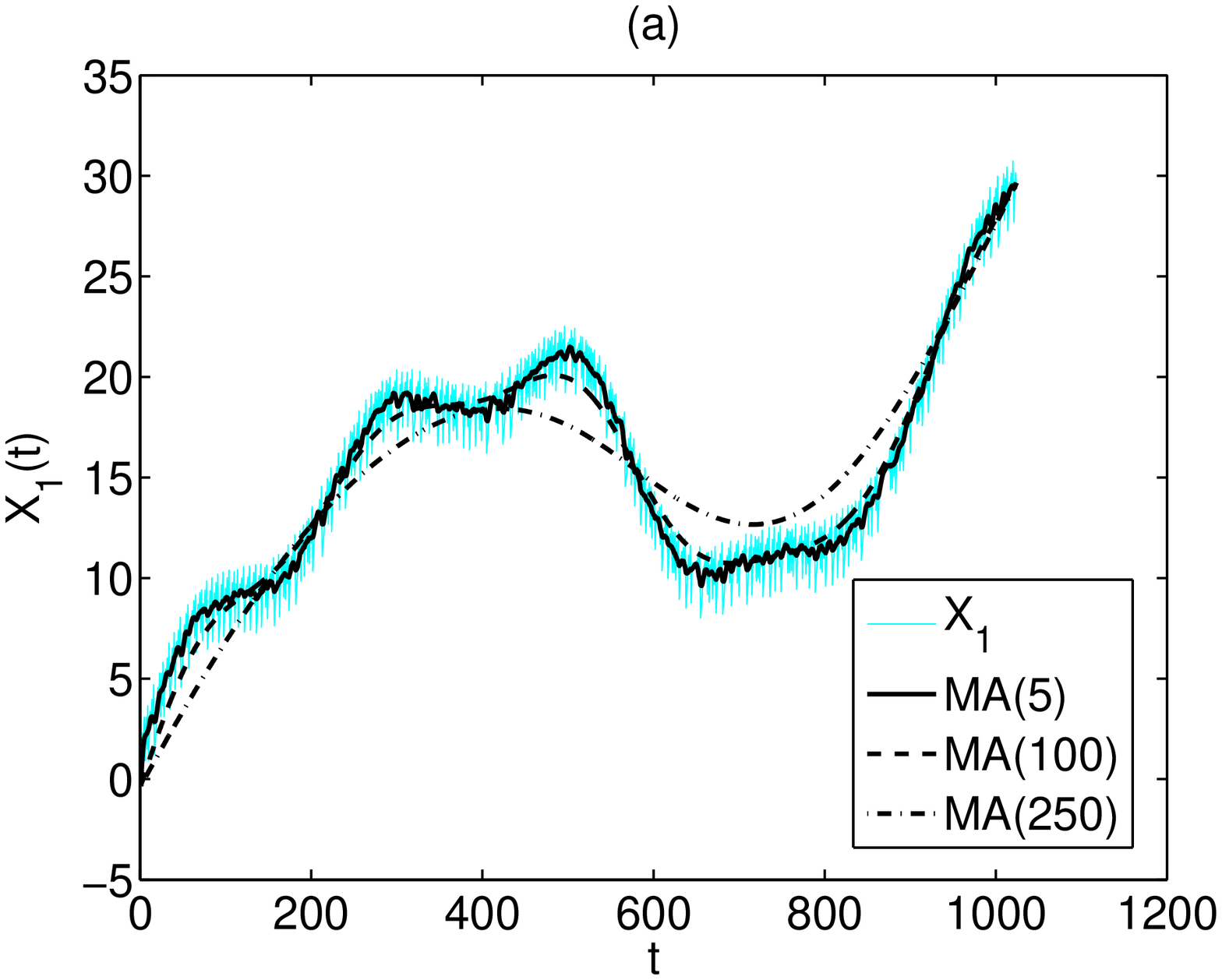}
\includegraphics[width=6cm]{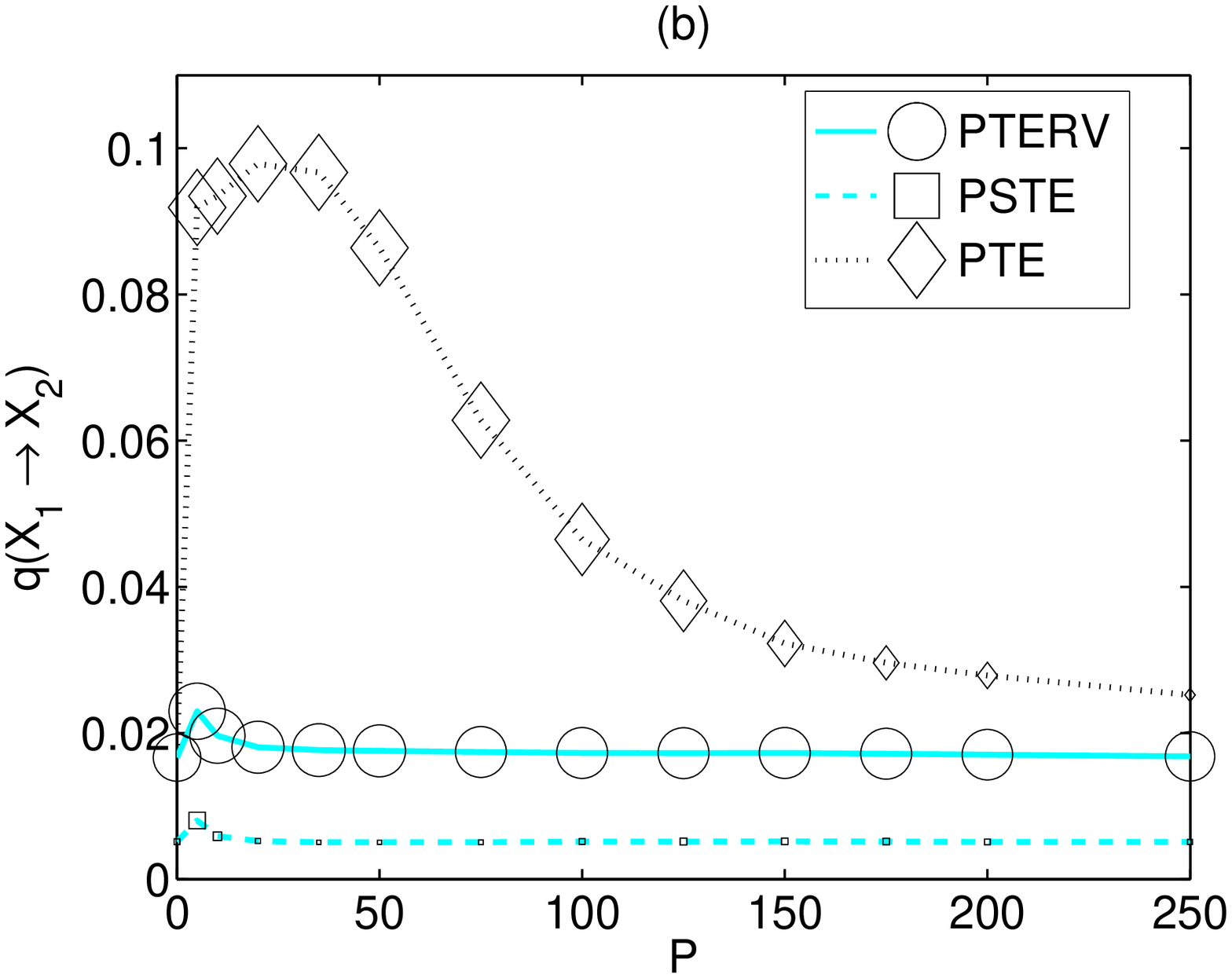}}
\includegraphics[width=6cm]{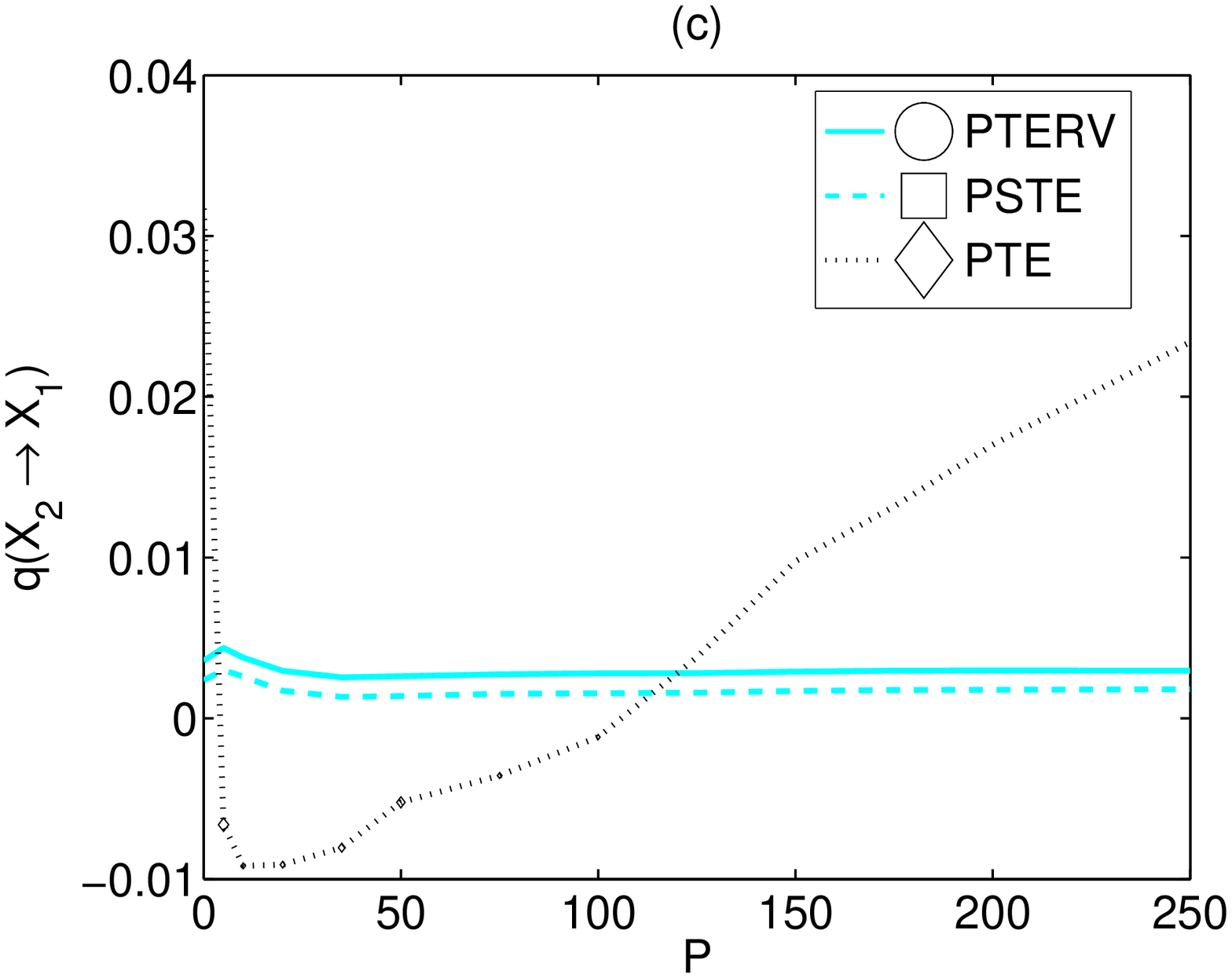}
\caption{(a) A time series of the variable $X_1$ of the coupled H\'{e}non maps ($K=3$) after adding to it a stochastic trend and superimposed is the moving average smoothing of different orders $P$, as shown in the legend. (b) and (c) PTERV, PSTE and PTE measures (denoted collectively $q$) as functions of the order $P$ of the moving average smoothing used to detrend each time series from the coupled H\'{e}non maps of 3 variables ($N=1024$, $C=0.2$, $m=2$, $\tau=1$, $T=1$). The number of rejections in 100 realizations of the randomization test determines the size of a symbol displayed for each measure and $P$, where in the legend the size of the symbols regards 100 rejections. The panel in (b) is for the coupling $X_1 \rightarrow X_2$ and in (c) for $X_2 \rightarrow X_1$. }
\label{fig:diffMAorders}       
\end{figure}

The type of detrending turns out to have very little effect on the rank measures but it has on PTE. For the same example, and a weaker stochastic trend with the SD of the random walk increments being 0.6 times the SD of the coupled H\'{e}non time series, we apply smoothing with moving average of different orders $P$. As shown in Fig.~\ref{fig:diffMAorders}a, a small $P$ (but not as small to wash out the true dynamics, say one or two) overfits the stochastic trend while a large $P$ leaves some drift in the time series. 
A best $P$ is hard to find, and even when using the same $P$ as for the generation of the stochastic trend ($P=100$), some small drift can be seen to remain, knowing that the true time series does not contain any slow fluctuations. We computed PTERV, PSTE and PTE and made randomization tests on 100 realizations for a range of orders $P$ of moving average detrending, and the results for the true direct coupling $X_1 \rightarrow X_2$ and the non-existing coupling $X_2 \rightarrow X_1$ are shown in Fig.~\ref{fig:diffMAorders}b and c, respectively. PTERV has striking robustness to the level of detrending and both its value and the percentage of rejections for the randomization test are at the same level for any $P$, including $P=0$ that regards no detrending. PSTE is equal stable but fails to detect the true direct coupling as was the case with stationary time series. PTE is affected by $P$ and becomes very irregular. For the true coupling $X_1 \rightarrow X_2$, PTE increases with $P$ for small $P$ up to a peak and then decreases with $P$. Accordingly, the power of the randomization test decreases down to the zero level for $P=250$ reaching the same performance as for the generated non-stationary time series. Remarkably, for the non-existing coupling $X_2 \rightarrow X_1$, the pattern of the dependence of PTE on $P$ is the opposite, and for the range of $P$ the true coupling was best detected, a larger number of rejections than the nominal significance level is observed for the randomization test. These results show that PTERV, unlike PTE, can directly be applied to multivariate time series with trends and avoid thus detrending that may alter the dynamics and inter-dependencies.

\section{Discussion}
\label{sec:Discussion}

The rank measures of causality of partial transfer entropy on rank vectors (PTERV) and partial symbolic transfer entropy (PSTE) extend the bivariate causality measures TERV and STE, respectively, to account for the presence of other observed confounding variables. In the same way that TERV was developed to modify the future response vector in STE in order to correspond exactly to transfer entropy (TE), PTERV is the direct analogue of partial transfer entropy (PTE), substituting samples with ranks. PTERV has a clear advantage over PTE if the latter is estimated by binning methods because the domain of joint probabilities for PTERV is generally much smaller (for a dimension $m$ the size of the domain for ranks is $m!$ and for the binning is $b^m$ where $b$ is the number of bins). Therefore in our simulation study we compared PTERV and PSTE with PTE using an advanced estimate of nearest neighbors that is found to be much more efficient than any binning estimate. 
  
We attempted to determine the asymptotic properties of PTERV, and subsequently PSTE, approximating first the bias and variance, and then the distribution of PTERV when there is no coupling. We considered approximations with Gaussian and Gamma distributions expressing their parameters in terms of the estimated bias and variance, and also using known results on Gamma approximation of conditional mutual information. We considered these approximations for the null distribution of parametric tests for the significance of PTERV (and PSTE). However, for the problem of assessing direct coupling, there are often certain dependence structures of the driving and the response variable on the confounding variables, which the analytic approximations do not encompass and thus the parametric tests are likely to fail. We confirmed this with simulations that showed also that the randomization test using time-shifted surrogates, though it cannot encompass all types of dependencies, is superior to the parametric tests. More work is needed to improve the analytic approximation of the distribution of PTERV, and direct causality measures in general, so as to make the parametric tests more accurate. 

The correct detection of the true direct causality is a difficult task, and depending on the underlying intrinsic dynamics of each subsystem and the coupling structure of them, one has to search for the optimal setting of the reconstruction parameters for all involved variables (response, driving, conditioning) as well as the horizon of the time ahead. The rank measures were found to be particularly sensitive to the choice of the reconstruction parameters. The embedding dimension has significant effect on the performance of the rank causality measures in the same way that the estimation of the probability mass function depend heavily on the dimension of the discrete variable. Moreover, while PTERV can estimate better direct causal effects than PSTE for small $m$, as $m$ increases the two measures converge because the difference in the definition of the future response vector is small in the context of large embedding vectors. The simulations on chaotic coupled systems showed that PTERV performs always better than or equally good as PSTE. Compared to PTE using the nearest neighbor estimate, PTERV is more data demanding and thus the power of the randomization test using PTERV is smaller. 

PTERV is robust to drifts in the time series and the simulations showed that both the power and significance of the randomization test with PTERV remains stable when stochastic trend is added in the time series, as well as under different levels of detrending. For these data conditions, PTE is found inappropriate and the same holds for any causality measure based on samples rather than ranks.  

\bibliographystyle{epj}


\end{document}